\documentclass[12pt,twocolumn]{emulateapj_Astroph}

\tightenlines

\usepackage{ulem} 
\usepackage{color} 
\usepackage{epsfig,amsmath}
\usepackage{times}

\newcount\listnorom
\newcommand\listromanDE{\global\advance \listnorom by 1
{\lowercase\expandafter{(\romannumeral\listnorom)}\ }}
\newcommand\newlistroman{\listnorom=0}

\def\lsim{\raise0.3ex
  \hbox{$<$\kern-0.75em\raise-1.1ex\hbox{$\sim$}}\,}
\def\gsim{\raise0.3ex
  \hbox{$>$\kern-0.75em\raise-1.1ex\hbox{$\sim$}}\,}

\newcommand{\Unew}{u_\mathrm{new}}
\newcommand{\sigKW}{\sigma_\mathrm{KW}}
\newcommand{\CenA}{Centaurus~A}
\newcommand{\GZK}{Greisen-Zatsepin-Kuzmin}
\newcommand{\Xsub}{x_\mathrm{sub}}
\newcommand{\DelFen}{\Delta F_E}
\newcommand{\DelFp}{\Delta F_p}

\newcommand{\LAS}{large-angle-scattering}

\newcommand{\Ng}{N_g}

\newcommand{\gyrotime}{\tau_g}
\newcommand{\delMax}{\delta\theta_\mathrm{max}}
\newcommand{\Tscat}{\theta_\mathrm{scatt}}
\newcommand{\delAng}{\delta\theta}
\newcommand{\GamZ}{\Gamma_0}
\newcommand{\GamTwo}{\Gamma_2}
\newcommand{\gamRS}{\gamma_\mathrm{RS}}
\newcommand{\gamEj}{\gamma_\mathrm{ej}}
\newcommand{\GamTwoMC}{\Gamma_\mathrm{2,MC}}

\newcommand{\gamZ}{\gamma_0}
\newcommand{\betaZ}{\beta_0}
\newcommand{\gamrel}{\gamma_\mathrm{rel}}
\newcommand{\betarel}{\beta_\mathrm{rel}}

\newcommand{\vHT}{v_\mathrm{HT}}
\newcommand{\HT}{\hbox{de~Hoffmann}-�Teller}
\newcommand{\RH}{Rankine-Hugoniot}

\newcommand{\rg}{r_g}

\newcommand{\pmax}{p_\mathrm{max}}
\newcommand{\Pesc}{p_\mathrm{esc}}
\newcommand{\Eesc}{E_\mathrm{esc}}

\newcommand{\QescPx}{Q_\mathrm{px}^{\mathrm{FEB}}}
\newcommand{\QescPxMC}{Q_\mathrm{px,MC}^{\mathrm{FEB}}}
\newcommand{\QescEn}{Q_\mathrm{en}^{\mathrm{FEB}}}
\newcommand{\QescEnMC}{Q_\mathrm{en,MC}^{\mathrm{FEB}}}
\newcommand{\qescEn}{q_\mathrm{en}^{\mathrm{FEB}}}

\newcommand{\pcc}{cm$^{-3}$}

\newcommand{\EffDSA}{{\cal E_\mathrm{DSA}}}

\newcommand{\Pmc}{P_\mathrm{MC}}
\newcommand{\Pxx}{P_\mathrm{xx}}
\newcommand{\SA}{semi-analytic}

\newcommand{\xx}[1]{\!\times\!10^{#1}}

\newcommand{\Fnum}{F_\mathrm{n}}
\newcommand{\FnumZ}{F_\mathrm{n}^0}
\newcommand{\Fen}{F_\mathrm{en}}
\newcommand{\FenZ}{F_\mathrm{en}^0}
\newcommand{\Fpx}{F_\mathrm{px}}
\newcommand{\FpxZ}{F_\mathrm{px}^0}

\newcommand{\DSA}{diffusive shock acceleration}

\newcommand{\MFA}{magnetic field amplification}
\newcommand{\rgz}{r_{g0}}
\newcommand{\rgZ}{r_{g0}}

\newcommand{\kmps}{km~s$^{-1}$}
\newcommand{\NLin}{Nonlinear}
\newcommand{\NL}{nonlinear}

\newcommand{\UuM}{Unmodified}

\newcommand{\Lfeb}{L_\mathrm{FEB}}

\newcommand{\alf}{Alfv\'en}

\newcommand{\SC}{self-consistent}
\newcommand{\SCly}{self-consistently}
\newcommand{\rNL}{r_\mathrm{NL}}
\newcommand{\Rtot}{r_\mathrm{tot}}

\newcommand{\rRH}{r_\mathrm{RH}}
\newcommand{\muG}{$\mu$G}

\newcommand{\Emax}{E_\mathrm{max}}

\newcommand{\be}{\begin{eqnarray}}
\newcommand{\ee}{\end{eqnarray}}

\newcommand{\Lmfp}{\lambda_\mathrm{mfp}}
\newcommand{\etamfp}{\eta_\mathrm{mfp}}

\newcommand{\rel}{relativistic}
\newcommand{\Rel}{Relativistic}
\newcommand{\nonrel}{non\-rel\-a\-tiv\-is\-tic}

\newcommand{\transrel}{trans-rel\-a\-tiv\-is\-tic}
\newcommand{\Transrel}{Trans-rel\-a\-tiv\-is\-tic}
\newcommand{\ultrarel}{ul\-tra-rel\-a\-tiv\-is\-tic}
\newcommand{\Ultrarel}{Ul\-tra-rel\-a\-tiv\-is\-tic}
\newcommand{\degg}{^\circ}
\newcommand{\Tbn}{\theta_\mathrm{Bn}}

\newcommand{\Tht}{\theta_\mathrm{HT}}
\newcommand{\TP}{test-particle}
\newcommand{\Unmod}{unmodified}
\newcommand{\PA}{pitch-angle}
\newcommand{\PAS}{pitch-angle scattering}
\newcommand{\deltime}{\delta t}

\newcommand{\mc}{Monte Carlo}
\newcommand{\MC}{Monte Carlo}
\newcommand{\syn}{synchrotron}

\newcount\Inum
\newcount\IInum
\newcount\IIInum
\newcount\IVnum
\def\I{\global\multiply\IInum by 0 \global\multiply\IIInum by 0
            \global\multiply\IVnum by 0 \global\advance \Inum by 1
            {\the\Inum. }}
\def\II{\global\multiply\IIInum by 0\global\multiply\IVnum by 0
       \global\advance \IInum by 1 {\the\Inum.\the\IInum. }}
\def\III{\global\multiply\IVnum by 0\global\advance \IIInum by 1
            {\the\Inum.\the\IInum.\the\IIInum. }}
\def\IV{\global\advance \IVnum by 1
            {\the\IVnum. }}
\begin{document}

\title{Monte Carlo simulations of nonlinear particle acceleration in parallel trans-relativistic shocks} 

\author{Donald C. Ellison,\altaffilmark{1},
Donald C. Warren\altaffilmark{1} and Andrei M. Bykov
\altaffilmark{2}}

\altaffiltext{1}{Physics Department, North Carolina State
University, Box 8202, Raleigh, NC 27695, U.S.A.;
don\_ellison@ncsu.edu}

\altaffiltext{2}{Ioffe Institute for Physics and Technology, 194021
St. Petersburg, Russia; ambykov@yahoo.com}

\begin{abstract}
We present results from a Monte Carlo simulation of a parallel
collisionless shock undergoing particle acceleration.
Our simulation, which contains parameterized scattering and
a particular thermal leakage injection model, 
calculates the 
feedback between accelerated particles ahead of the shock, which 
influence the shock precursor and ``smooth"  the shock, and thermal particle injection.
We show that there is 
a transition between \nonrel\ shocks, where the acceleration 
efficiency can be extremely high and the 
\NL\ compression ratio can be substantially  greater than the \RH\ value, and fully \rel\ shocks, where \DSA\  is less efficient and the compression ratio remains at the  \RH\ value. 
This transition occurs in the \transrel\ regime and, for the particular parameters we use, occurs around a shock 
Lorentz factor $\gamZ =1.5$.
We also find that 
\NL\ shock smoothing dramatically reduces the acceleration efficiency presumed to occur with large-angle scattering in \ultrarel\ shocks.
Our ability to seamlessly treat the transition from \ultrarel\ to \transrel\ to \nonrel\ shocks 
may be important for evolving \rel\ systems, such as     gamma-ray bursts and type Ibc supernovae.
We expect a substantial evolution of shock accelerated spectra during this transition from soft early on to much harder when the 
blast-wave shock becomes \nonrel.
\end{abstract}

\keywords{ acceleration of particles, shock waves, ISM: cosmic rays,
           ISM: supernova remnants, gamma-ray bursts}

\section{Introduction}
Collisionless shocks are known to accelerate particles, and \DSA\ 
(DSA) ( also known as the first-order Fermi mechanism) is widely regarded as the most likely mechanism for this acceleration 
\citep[e.g.,][]{Kry77,ALS77,Bell78a,BO78,JE91,MD2001,
BykovEtal2012, SchureEtal2012}.
Most shocks of astrophysical interest are \nonrel, where the speed of the shock $u_0$ is a small fraction of the speed of light $c$. Relativistic shocks, where the flow speed Lorentz factor $\gamZ=[1 -(u_0/c)^2]^{-1/2}$ is significantly greater than unity, are both less common and more difficult to study analytically. 
Because of these differences, and the fact that \nonrel\ shocks are both ubiquitous  and known to be efficient accelerators 
\citep[e.g.,][]{EMP90,WarrenEtal2005}, most work on DSA has been directed toward understanding  \nonrel\ shocks. 

There are some objects, however, where relativistic shocks are likely to be important in accelerating particles
\citep[see][]{BykovTreumann2011,BykovEtal2012}. These include gamma-ray bursts (GRBs), \rel\ supernovae (SNe), pulsar wind nebulae (PWNe), extra-galactic radio jets, and blazars. Relativistic shocks are more difficult to describe analytically 
because the assumption of particle isotropy breaks down. As long as $u_{0}$ is much less than the individual particle speed $v_p$, one can make the diffusion approximation that particles are roughly isotropic in all frames. As the shock becomes relativistic, $v_p \sim u_{0} \sim c$ and the diffusion approximation no longer applies 
\citep[see, however,][ and references therein for analytic approaches to the problem of relativistic shocks]{KirkEtal2000,BV2005}.

To date the most fruitful approaches to DSA by relativistic shocks have been particle-in-cell (PIC) and Monte Carlo simulations. PIC simulations 
\citep[e.g.,][]{GE2000,NishikawaEtal2007,Spitkovsky2008b,SironiSpit2011} can directly model not only the particle acceleration process but the shock formation process as well. They have a great advantage over analytic and \mc\ techniques 
\citep[e.g.,][]{Ostrowski1988,EJR90,ED2002,LP2003,Baring2004,NO2006}
in that they determine the self-generated magnetic turbulence, and can treat ions and electrons, \SCly. As is well known, 
details of the wave-particle interactions are far more important in \rel\ shocks than in \nonrel\ ones, so determining the magnetic turbulence \SCly\ is critical 
\citep[e.g.,][]{PLM2009,PPL2013}. The biggest disadvantage of PIC simulations is that they are extremely computationally intensive and must be fully three dimensional in order to model the full effects of cross-field diffusion \citep*[see][ for a derivation of the restricted dimensionality constraint]{JJB98}.

The \mc\ technique used in this paper treats particle scattering and transport explicitly, which, in effect, solves the Boltzmann equation with collective scattering \citep[e.g.,][]{EE84}. 
It contains a specific thermal injection model whereby some fraction of shock-heated particles are able to re-cross the shock and gain additional energy.
This injection process is a direct result of the scattering assumptions 
and requires no additional parameters.

Once the scattering assumptions are made (as detailed in 
Section~\ref{sec:diff} below), the 
shock structure, overall compression ratio $\Rtot \equiv u_0/u_2$, and full particle distribution function $f(x,p)$ 
at all positions relative to the subshock, are obtained 
self-consistently. 
Here and elsewhere the subscript ``0" implies far upstream values and the subscript ``2" implies downstream values, i.e., $x>0$.

In terms of applicability, the \mc\ model  lies between analytic techniques and PIC simulations. 
Analytic techniques are computationally faster but they have difficulty treating thermal particle injection.
On the other hand, our \MC\ calculations are much faster than PIC simulations and
they can provide important results at all shock speeds, including \transrel\ flows.
The main disadvantages of our \mc\ technique are that it is intrinsically steady state and plane-parallel, it is computationally intensive compared to \SA\ techniques, 
and it assumes a form for the scattering operator and so does not model the magnetic turbulence generation or the shock 
formation process \SCly.\footnote{We note that \nonrel\ versions of our \MC\ code can model magnetic turbulence generation in a fairly consistent manner \citep*[e.g.,][]{VBE2008,VBE2009}.}

While we have emphasized some 
differences between \nonrel\ and \rel\ shocks, all shocks must conserve momentum and energy and
\Unmod\ (UM) shocks undergoing efficient particle acceleration do not 
\citep[e.g.,][]{Drury83,BE87}.\footnote{An UM shock is one in which the plasma flow speed drops discontinuously at the shock front from 
the far upstream value $u_0$ to the downstream value $u_2$, both measured in the shock rest frame.}
If a shock, regardless of its speed, puts a non-negligible amount of energy into superthermal particles via DSA, the backpressure of the accelerated particles slows the upstream plasma before it reaches the discontinuous subshock (see Figure~\ref{fig:profile_gam1.5} in 
Section~\ref{sec:fine_scat}). This effect is \NL\ in nature, and acts to modify the shock structure in order to conserve momentum and energy.
Apart from PIC results, and the preliminary work of \citet{ED2002}, we know of no \SC, \NL, \rel\ and \transrel\ shock solutions other than those presented here.

The mathematical and computational difficulties in treating NL effects notwithstanding, \rel\ shocks are expected to be intrinsically less efficient than \nonrel\ ones, making NL effects less important.\footnote{This stems mainly from the smaller compression 
ratio ($\Rtot \sim 4$ for strong \nonrel\ shocks versus $\Rtot \sim 3$ for \ultrarel\ shocks). In addition there is more uncertainty concerning the ability of \rel\ shocks to self-generate the magnetic turbulence that is necessary for DSA to 
occur \citep*[e.g.,][]{PLM2009}. Acceleration beyond the initial Lorentz boost remains highly uncertain in \ultrarel\ shocks.} 
While this may be the case, we show that NL effects can be important in \rel\ shocks depending on the particular assumptions made regarding particle diffusion. Specifically, if \LAS\ (LAS) occurs, where the magnetic turbulence is strong enough to randomize the particle trajectory in just a few interactions, NL effects can be dramatic and cannot be ignored in any realistic astrophysical application.

In this paper we expand on previous \MC\ 
calculations \citep[i.e.,][]{ED2002,ED2004,DoubleEtal2004}. 
We include  particle escape at an upstream 
free escape boundary (FEB) and emphasize
\transrel\ shocks. \Transrel\ shocks are likely to be important for GRB afterglows, where CR production, perhaps to the highest observed energies \citep*[e.g.,][]{KMW2010}, may occur. 
A sub-population of type Ibc supernovae, not producing observed GRBs, have observed \transrel\ speeds and these have been proposed as sources of ultra-high-energy cosmic rays (UHECRs)
\citep[e.g.,][]{Soderberg2010,Chakraborti2011}.
Of course, modeling astrophysical observations normally means modeling electrons since they radiate far more efficiently than do ions. 
While electrons can be included in our \MC\ simulation
\citep[e.g.,][]{BaringEtal99}, doing so requires additional ad hoc
assumptions. We do not attempt a combined treatment here, deferring the issue to a future paper.

The application of DSA to pulsar winds brings another important difficulty to the forefront: the shock obliquity, that is, the angle between the shock normal and the upstream
magnetic field, $\Tbn$. \Rel\ shocks are certain to be highly oblique since the downstream magnetic field is bent away from the downstream shock normal more strongly than in \nonrel\ shocks \citep*[see, for example,][]{ED2004,Meli_Quenby2008}. This effect increases with the shock Lorentz factor so the downstream angle in \ultrarel\ pulsar wind termination shocks should be almost perpendicular.
While shocks of any obliquity can be treated directly with PIC  simulations if they are done in 3D, oblique geometry makes 
\SA\ and \MC\ models more difficult. \UuM, oblique, \rel\ shocks have been discussed extensively 
\citep[e.g.,][]{Ostrowski1991,ED2004,LR2006,NiemiecEtal2006,SummerlinBaring2012}
but we know of no non-PIC  
treatments of \SC, \rel, oblique shocks. Here, we confine our discussion to parallel shocks where $\Tbn=0\degg$
and plan to treat modified oblique shocks in future work.

\newlistroman

Our NL parallel shock simulations show several important features: 
\listromanDE shock modification results in significant changes to  the accelerated particle distribution for all Lorentz factors including fully \rel\ shocks, even though they are intrinsically less efficient accelerators than \nonrel\ shocks;
\listromanDE
\rel\ shocks with coarse, LAS attempt to accelerate particles efficiently, but this produces strong shock modification that dramatically  softens the accelerated particle spectrum; 
\listromanDE
in going from fully \rel\ to \nonrel\ shock speeds, as might be relevant for modeling \transrel\ supernovae and
GRB afterglows, there is a relatively
sharp transition from soft spectrum \rel\ shocks (i.e., phase-space
spectra $f(p) \propto p^{-4.23}$) to  \transrel\ 
shocks with $\Rtot > 4$ and concave upward
spectra typical of efficient \nonrel\ shocks; and
\listromanDE
in \transrel\ shocks, a lowering of the ratio of specific heats can result in compression ratios greater than the \RH\ value without particle escape.  

The structure of the remainder of this paper is as follows. 
In Section~\ref{sec:Model}, we discuss relevant aspects of the \mc\ approach. We present results from our simulations in 
Section~\ref{sec:Results}, paying special attention to \transrel\ cases. 
In Section~\ref{sec:Diss}, we discuss these results in the context of astrophysical settings like GRBs, and we conclude in 
Section~\ref{sec:Conc}.

\section{Model} \label{sec:Model}
\subsection{Particle Diffusion}\label{sec:diff}
The details of the scattering process we use are described in 
\citet*{EJR90,EBJ96} 
and \citet{ED2002,ED2004}. Very generally \citep[and paraphrasing the discussion in][]{ED2004}, the scattering properties of the medium are modeled with the two parameters $\etamfp$ and $\Ng$.

We assume that particles \PA\ scatter with a mean free path
\begin{equation} \label{eq:mfp}
\Lmfp = \etamfp \rg
\ ,
\end{equation}
where $\rg=pc/(eB_0)$ is the gyroradius and $\etamfp$ parameterizes the ``strength" of scattering. Here, $p$ is the particle momentum in the local frame, $c$ is the speed of light, $e$ is the electronic charge, and $B_0$ is the uniform magnetic field in Gauss used to scale $\rg$.
Note that we do not attempt to model 
self-generated turbulence or \MFA\ 
\citep[e.g.,][]{MV82,Bell2004,Bell2005,VBE2008}. 
Nor do we trace particle orbits in a magnetized background 
\citep[e.g.,][]{NiemiecEtal2006,LR2006}.
Acknowledging that particle scattering in \rel\ shocks is more complex, we  
simply 
assume magnetic turbulence exists throughout the shock such that equation~(\ref{eq:mfp}) holds independent of position.

The strong scattering limit of $\etamfp=1$ is called Bohm diffusion and is commonly assumed because of its simplicity. There is little direct evidence for Bohm diffusion in actual sites where DSA takes place.
\citet{Uchiyama_J1713_2007} examined the variability of X-ray bright knots behind the \nonrel\ shocks of RXJ1713.7-3946; they concluded that the short timescales on which the knots dimmed and brightened require $\etamfp\sim 1$.
Using X-ray and GeV afterglows in GRBs, \citet{SagiNakar2012} have suggested that acceleration can approach the Bohm limit when the blast wave is  
\ultrarel. Nevertheless, these estimates are indirect and highly uncertain and some 
simulations of \rel\ and \transrel\ shocks \citep[e.g.,][]{LR2006} imply that $\etamfp\gsim10$ or higher. Furthermore, it has been known for some time that high-energy particles scattering in small-scale turbulence may have $\Lmfp \propto p^2$ rather than 
$\Lmfp \propto p$ \citep[e.g.,][]{Jokipi1971,PPL2011}.

In parallel shocks, $\etamfp$ sets the length and time scales for acceleration and $\etamfp \gg 1$ may severely  limit the maximum CR energy a given shock can produce. In oblique shocks, $\etamfp \gg 1$ will reduce the injection efficiency \citep[see][]{ED2004}.
For our examples here, we use $\etamfp=1$. However, 
in our parallel-shock approximation,
 larger values of $\etamfp$ only increase the  length and time scales of the simulation. The only length scale present in our steady-state model is the FEB. We have confirmed that simulations where the FEB is fixed in diffusion lengths give results independent of $\etamfp$.

For \PA\ scattering, if the time in the local frame required for the particle momentum vector to accumulate deflections of the order of $90\degg$ is identified with the collision time $t_c = \Lmfp/v_p$, where $v_p$ is the particle speed in the local frame, it was shown in the above references that 
the maximum deflection a particle experiences in an interaction satisfies
\begin{equation} \label{eq:delMax}
\delMax = \sqrt{6 \deltime/t_c}
\ ,
\end{equation}
where $\deltime$ is the time in the local frame between \PA\ scatterings.
The stochastic scattering process is simulated with two random numbers, $\phi$ and $\delAng$. Referring to Figure~2 in \citet{ED2004}, at each time step $\deltime$, the azimuthal angle $\phi$ is chosen randomly between 0 and $2\pi$ and the particle pitch angle, $\delAng$, is chosen from a uniform distribution of $\cos{\delAng}$ between  1 and $\cos{\delMax}$.

The maximum pitch angle deflection $\delMax$ is determined by setting $\deltime = \gyrotime/\Ng$, where $\Ng$ 
is the number of 
gyro-time segments $\deltime$ dividing a gyro-period $\gyrotime=2\pi\rg/v_p$, yielding
\begin{equation}\label{eq:Tmax}
\delMax = \sqrt{12 \pi /(\etamfp \Ng)}
\ .
\end{equation}
The parameter $\Ng$ determines the 
``fineness" of scattering, i.e., how many \PA\ scatterings on average, each with maximum deflection $\delMax$, are required to scatter the particle through $\sim \! 90\degg$.
Small values of $\Ng$ imply ``large-angle scattering" (LAS) where the particle direction is randomized in a few interactions with the background turbulence.
The size of $\delMax$ also depends on $\etamfp$. For example, a large $\etamfp$ implies weak scattering, with only a small deflection at any particular interaction. 
Note that the specific
value of $\etamfp$ implies that magnetic fluctuations with correlation lengths on the order of $L_c = 2 \pi \etamfp \rg /\Ng$ exist with sufficient power to produce this scattering.
The LAS regime is unlikely to be relevent for \ultrarel\ shocks but may occur in \transrel\ shocks \citep[see][ for a discussion of turbulence generation in \rel\ shocks]{PLM2006}.

We further assume  that all scatterings are elastic and isotropic in the frame where the scattering occurs
and that the scattering centers are frozen in the fluid, thus eliminating the ability to model second-order Fermi acceleration or the transfer of energy between particles and magnetic field via the production or damping of magnetic turbulence. We also neglect any cross-shock electric potential that may exist. 
The \mc\ model does \SCly\ include so-called shock-drift acceleration although this process, whereby particles gain energy as they gyrate in a compressed magnetic field, is not important in the parallel-shock results we show here. 
Again we emphasize the intermediate nature of the \mc\ 
technique: while it is clear that the approximations we employ to model the wave-particle interactions are severe compared to the \SC\ nature of PIC simulations, or even \MC\ simulations that trace particle orbits in prescribed background turbulence, they are more general than what is currently possible with \SA\ solutions. Given our parameterization of the plasma interactions we can model shock smoothing and NL acceleration over a dynamic range well beyond that currently accessible with 3-D PIC simulations. 

\subsection{Shock Modification and Particle Escape}\label{sec:Mod}
The \SC\ shock structure is found with a method
that iterates on three quantities: the flow speed profile $u(x)$, the overall shock compression ratio $\Rtot = u_0/u_2$, and the subshock length scale, $\Xsub$ 
\citep*[see][ for more details]{EBJ96}.
An initial $\Rtot$, $u(x)$, and $\Xsub$ 
are chosen, the simulation is run and the momentum and energy fluxes are calculated at all $x$-positions.
If the fluxes are not conserved, i.e., if they are not equal at all $x$-positions to the far upstream values, a new profile $u(x)$ is chosen and the simulation repeated with the same $\Rtot$. 

At each iteration where the fluxes are not conserved, we predict the next $u(x)$, i.e., $\Unew(x)$, using the fully \rel, steady-state, \RH\ conservation relations in the parallel shock frame
\citep[see, for example,][]{DoubleEtal2004}:
\begin{equation} \label{eq:NumFlux}
\Fnum(x) = \gamma(x) n(x) \beta(x)  = \FnumZ 
\ ,
\end{equation}
\begin{eqnarray} \label{eq:PxFlux}
\Fpx(x) &=& 
\gamma^2(x) \beta_x^2(x) [e(x) + \Pxx(x)] + \Pxx(x) + \QescPx
\nonumber \\
&=& \FpxZ
\ ,
\end{eqnarray}
and
\begin{equation} \label{eq:EnFlux}
\Fen(x) = 
\gamma^2(x) \beta_x(x) [e(x) + \Pxx(x)] + \QescEn
= \FenZ 
\ .
\end{equation}
Here, $\FnumZ$, $\FpxZ$, and $\FenZ$ are the far upstream number, momentum, and energy fluxes, respectively, and we have explicitly included the escaping momentum and energy fluxes ($\QescPx$ and $\QescEn$) at an upstream free escape boundary (FEB).

Particle escape has been modeled extensively with either an upstream 
FEB \citep[e.g.,][]{EMP90} or a cutoff at some maximum momentum $\pmax$ \citep[e.g.,][]{EE84}. In either case, the highest energy particles leave the system and carry away energy and 
momentum flux.\footnote{Typically, the fraction of escaping particles is too small for the escaping number flux to be important.}
Here, we use an upstream FEB exclusively.

The fluxes $\FnumZ$, $\FpxZ$, and $\FenZ$ are determined from the inputted 
far upstream values, $\rho_0=m_p n_0$, $u_0$, $P_0$, and $\GamZ=5/3$.
Here $n_0$ is the proton number density, $P_0$ is the scalar
thermal pressure, and $\GamZ$ is the ratio of specific heats. 
In the above equations, $e$ is the total energy density and $\Pxx$ is the $xx$-component of the pressure tensor $\mathcal{P}$. 
Following 
\citet{DoubleEtal2004}, we write the adiabatic and gyrotropic equation of state as,
\begin{equation}\label{eq:EqState}
e(x) = \frac{P(x)}{\Gamma(x) -1} + \rho(x) c^2
\ ,
\end{equation}
where $\rho c^2$ is the rest mass energy density, $\Gamma$ is the adiabatic index, and $P=\Pxx$ is the isotropic scalar pressure, i.e., 
$P = Tr(\mathcal{P})/3$. 

We note that while $\Gamma$ is well defined for 
\nonrel\ ($\Gamma=5/3$) and \ultrarel\ particles ($\Gamma=4/3$), it depends on an unknown relation between $e$ and $P$ for \transrel\ particles.
In our iteration scheme, we use a simple scaling relation for $\Gamma(x)$ in 
equations~(\ref{eq:PxFlux}) and (\ref{eq:EnFlux})
\citep[see][]{DoubleEtal2004}:
\begin{equation}\label{eq:SpHeat}
\Gamma(x) = \frac{4 \gamrel(x) + 1}{3 \gamrel(x)}
\ ,
\end{equation}  
where
\begin{equation}
\gamrel(x) = [1 - \betarel(x)^2]^{-1/2}
\ ,
\end{equation}  
and 
\begin{equation}\label{eq:beta_rel}
\betarel(x) = \frac{\betaZ - \beta(x)}{1 - \betaZ\beta(x)}
\ .
\end{equation}
Here, $\betaZ=u_0/c$ and $\beta(x)=u(x)/c$.

After each iteration, we obtain from the simulation
\begin{equation}
\DelFp(x) = \FpxZ - [\Fpx(x) + \QescPxMC]
\ ,
\end{equation}
and
\begin{equation}
\DelFen(x) = \FenZ - [\Fen(x) + \QescEnMC]
\ ,
\end{equation}
that is, the differences between the known upstream fluxes 
and the \MC\ results.
The subscript `MC' on $\QescPxMC$ and $\QescEnMC$ indicates that these values are measured from the \MC\ simulation.
A consistent and unique solution is obtained, within statistical limits,
when  $\DelFp(x) \simeq \DelFen(x) \simeq 0$, $\QescPxMC \simeq \QescPx$, and $\QescEnMC \simeq \QescEn$.

If these conditions are not met, 
we obtain a new flow speed profile $Unew(x)$ as follows.
First, the pressure is obtained from equation~(\ref{eq:PxFlux}) using $\Fpx(x)$
calculated in the \MC\ simulation:
\begin{equation} \label{eq:Pmc}
\Pmc(x) = \frac{[\Fpx(x)/a_1] - 
\gamZ \betaZ \rho_0 c^2 }
{a_1 \Gamma' + 1/a_1}
\ ,
\end{equation}
where we have made the substitutions:
\begin{equation}
a_1 = \gamma(x) \beta(x)
\ ,
\end{equation}
\begin{equation}
\gamma(x) \beta(x) \rho(x) = \gamZ \betaZ \rho_0
\ ,
\end{equation}
and
\begin{equation} \label{eq:Gam}
\Gamma' = \Gamma(x)/[\Gamma(x) - 1] 
\ .
\end{equation}
This pressure is now used in equation~(\ref{eq:PxFlux}), with
$\Fpx(x)$ replaced by $\FpxZ$, to solve for the new speed profile $\Unew(x)$. 

The above scheme iterates on $u(x)$ but will not yield a correct solution (i.e., $\DelFp(x) \simeq \DelFen(x) \simeq 0$, etc.)
unless the compression ratio $\Rtot$ is consistent with 
equations~(\ref{eq:NumFlux})--(\ref{eq:EnFlux}).
The consistency of $\Rtot$ is determined as follows.

Equations~(\ref{eq:NumFlux})--(\ref{eq:EnFlux}) 
start with six unknown 
downstream values:
$\rho_2$, $u_2$, $P_2$, $\GamTwo$, $\QescPx$, and $\QescEn$.
We note that the escaping fluxes $\QescPx$ and $\QescEn$ contribute to the downstream side of equations~(\ref{eq:PxFlux}) and 
(\ref{eq:EnFlux}) even though the particles escape at an upstream FEB.
One unknown can be eliminated if the escaping particles are fully \rel, as we assume, since their energy and momentum are related by  $\Eesc=\Pesc c$, and $\QescEn = c \QescPx$, regardless of 
shock speed.\footnote{The escaping momentum flux depends on the $x$-component of momentum but for particles escaping at an upstream FEB, the momentum is directed nearly in the $-x$-direction so $\QescEn \simeq c \QescPx$.}
Since the \MC\ simulation is run by setting  the compression ratio $\Rtot = u_0/u_2$, another unknown is eliminated. 
Furthermore,  because the \MC\ simulation calculates the full distribution function  we can calculate the downstream ratio of specific heats as
$\GamTwoMC=1 + P_2/(e_2 - \rho_2c^2)$, where $P_2$ and $e_2$ are calculated directly from $f(x>0,p)$.
If we replace $\GamTwo$ with $\GamTwoMC$, we have three equations and three unknowns.\footnote{In principle, $\Gamma(x)$ in 
equation~(\ref{eq:Gam}) could be calculated directly from $f(x,p)$ as is $\GamTwoMC$. In practice, we find that equation~(\ref{eq:SpHeat}) is simpler and gives satisfactory results for calculating $\Pmc(x)$.}  

We now have a system of equations where $\Rtot$ and $u(x)$ are used to  predict  $\QescEn$ and $\QescPx$ from a shock 
undergoing DSA.
A totally consistent solution would yield 
$\DelFp(x)= \DelFen(x) = 0$ at all $x$,
$\QescPxMC = \QescPx$, and $\QescEnMC = \QescEn$. If these conditions are not satisfied to within a few percent, we vary $\Rtot$ and iterate on $u(x)$ again.

In some cases, particularly fully \rel\ shocks, the paired
iterations on $u(x)$ and $\Rtot$ do not result in a consistent solution and a third iteration on the subshock scale $\Xsub$ must be performed.
In these cases, the subshock speed is defined as 
\begin{equation} \label{eq:Xsub}
u(x) = u_2 - 
\left [ u_2 - u(\Xsub) \right ]
\frac{\tan^{-1}{x}}{\tan^{-1}{\Xsub}}
\end{equation}
over the range 
$-|\Xsub| < x < 0$, where $\Xsub < 0$.
This scaling is discussed for specific examples in 
Section~\ref{sec:Results}.

The combined iteration over $\Rtot$, $u(x)$, and $\Xsub$
yields consistent smooth-shock solutions over the full range of shock speeds.
We note that the \mc\ simulation uses
equations~(\ref{eq:NumFlux})--(\ref{eq:EqState}) as a consistency check but these equations are not used directly in propagating particles or calculating fluxes in the simulation. All fluxes are calculated directly in the shock frame from the individual particle kinematics. 

\subsubsection{Examples with no DSA} \label{sec:NoDSA}
\begin{figure}
\epsscale{1.05}
\plotone{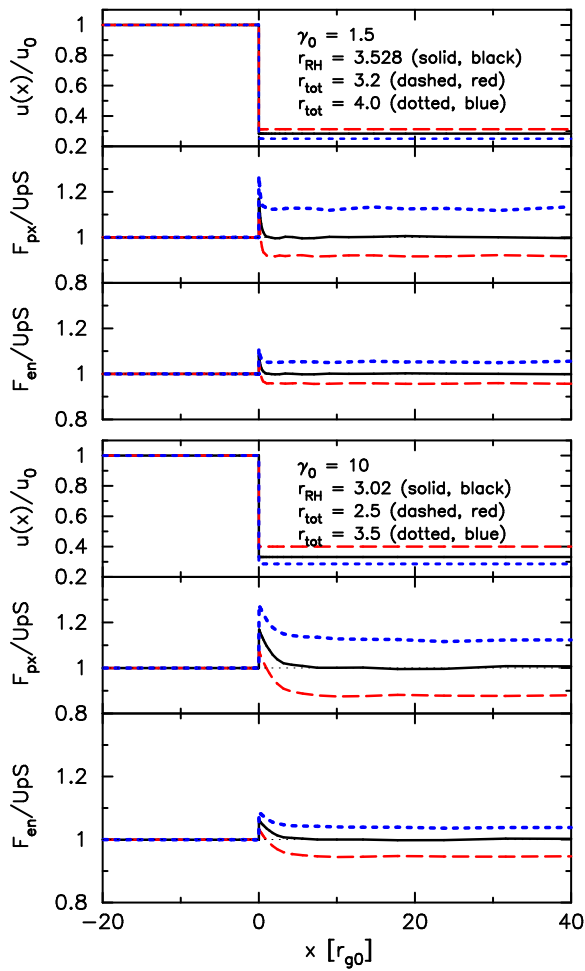} 
\caption{Shock rest frame values of the flow speed, $u(x)$, the momentum flux, $\Fpx$, and the energy flux, $\Fen$, are shown for UM shocks with no DSA.  All quantities are scaled to upstream values and the position $x$ is in units of the gyroradius 
$\rgz \equiv m_p u_0 c /(e B_0)$, where $B_0=3$\,\muG. 
The solid (black) curves assume a compression ratio $\Rtot = \rRH$, i.e., the \RH\ value as determined in \citet{DoubleEtal2004}. The 
dashed (red) and dotted (blue) curves are for lower and 
higher $\Rtot$'s as indicated.
\label{fig:comp_test}}
\end{figure}

In the absence of DSA (i.e., in an UM shock with 
$\QescPx = \QescEn=0$), 
equations~(\ref{eq:NumFlux})--(\ref{eq:beta_rel}) yield the \RH\ compression ratio $\rRH$, as determined by \citet{DoubleEtal2004}. Typically,  $\rRH$ is used to start our iteration process. 
In Figure~\ref{fig:comp_test} we show two examples with DSA disabled in the \MC\ simulation. The UM shock profiles are in the panels labeled $u(x)/u_0$ and the momentum and energy fluxes, scaled to far upstream values, are in the panels labeled $\Fpx/\mathrm{UpS}$ and $\Fen/\mathrm{UpS}$ respectively. For both shock Lorentz factors we show the fluxes for three compression ratios. The solid (black) curves are for $\rRH$
and the dashed (red) and dotted (blue) curves are for lower and higher $\Rtot$'s respectively.

We note that the upstream energy fluxes shown in 
Figure~\ref{fig:comp_test} and elsewhere in this paper differ 
from those in equation~(\ref{eq:EnFlux}) in that they ignore the incoming rest mass energy flux. 
While accelerated particles may carry away significant amounts of energy and momentum at a FEB, their density compared to the upstream plasma is always negligible, independent of almost all other factors. The rest mass energy flux entering the shock is therefore approximately
equal to the rest mass energy flux
leaving via downstream convection, no matter what the shock speed considered. Thus we ignore it and calculate escaping kinetic energy flux as a fraction of incoming kinetic energy 
flux.

Except for the region 
immediately downstream from the shock, the fluxes determined by the \mc\ simulation in Figure~\ref{fig:comp_test}
are precisely those determined by the \RH\ relations for $\rRH$, even for the \transrel\ case $\gamZ=1.5$. Results for compression ratios larger and smaller than $\rRH$ clearly do not conserve energy and momentum. 
The fact that the \MC\ results 
are consistent with $\rRH$ as determined from 
equations~(\ref{eq:NumFlux})--(\ref{eq:beta_rel}) is somewhat remarkable. 
As described in detail in \citet{ER91}, once the basic scattering assumptions given in 
Section~\ref{sec:diff} are made, a consistent shock solution  can be obtained with no further assumptions simply by running the simulation with different $\Rtot$'s.
There is no need to calculate a ratio of specific heats in the \MC\ simulation and the fluxes are calculated directly by summing over particles as they move with the bulk flow. The fluxes shown in 
Figure~\ref{fig:comp_test} are calculated directly in the shock frame and there is no need to transform into the local frame to calculate a pressure or energy density as required in 
equations~(\ref{eq:NumFlux})--(\ref{eq:beta_rel}).

Of course, the reason for defining 
equations~(\ref{eq:NumFlux})--(\ref{eq:EqState}) is that, with acceleration, $\Rtot$ can be greater than $\rRH$.
This occurs because  
$\GamTwo$ can be less than the \RH\ value and/or because
escaping particles carry away pressure causing $\Rtot > \rRH$
\citep[e.g.,][]{EE84,JE91,BE99}.
Both of these \NL\ processes are included \SCly\ in the \MC\ simulation.

\subsection{Particle Injection}\label{sec:inj}
While background superthermal particles will be injected and accelerated if they interact with a collisionless shock,
we discuss here only thermal leakage injection where all particles that get accelerated start as far upstream thermal particles.

\begin{figure}
\epsscale{1.0}
\plotone{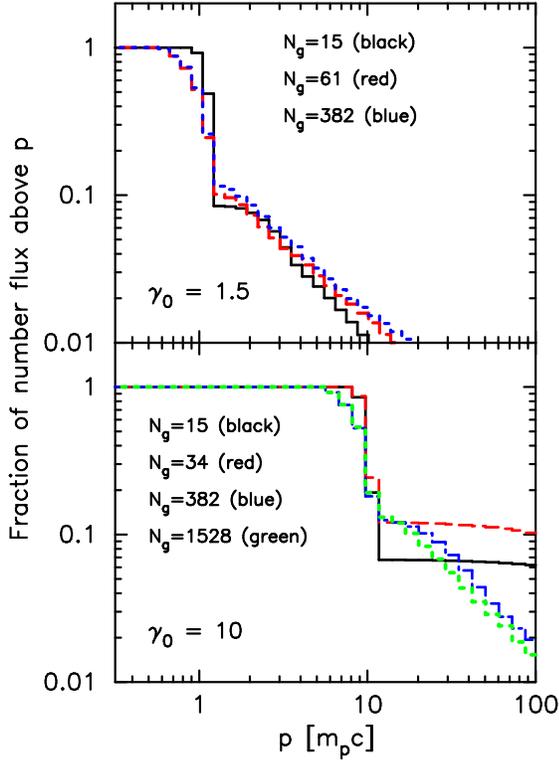} 
\caption{Fraction of proton number flux with shock frame momentum 
$p$ or greater versus $p$ for an UM shock with $\gamZ=1.5$ (upper panel) and 
$\gamZ=10$ (lower panel). Thermal particles enter the simulation far upstream from the shock and diffuse downstream. These spectra are calculated from particles that leave the shock far downstream. The values of $\Ng$ are related to the maximum pitch-angle scattering angle by equation~(\ref{eq:Tmax}). For $\Ng=15$, $\delMax \simeq \pi/2$.
\label{fig:Inj_frac}}
\end{figure}

In our \mc\ technique, thermal leakage injection is modeled in its simplest form. Unshocked thermal particles, injected far upstream, convect and diffuse with a mean free path in the local frame given by 
equation~(\ref{eq:mfp}). No restriction on $v_p/u$ is imposed.
At every pitch-angle scattering event a Lorentz transformation is performed such that the particle makes an isotropic and elastic scattering in the frame of the scattering interaction.
After crossing into the downstream region and interacting with the downstream plasma, some fraction of the particles will have a speed  $v_p > u_2$. These particles have a certain probability to diffuse back upstream and be further accelerated. 
In \rel\ shocks, this probability for returning back across the shock depends on the scattering parameters $\Ng$ and  $\Lmfp$\footnote{For an isotropic distribution of particles, it is well-known that the probability of return depends only on the downstream flow speed and an individual particle's speed 
\citep[see, for example,][]{Bell78a,EBJ96}. Near the shock, however, the distribution is highly anisotropic, and so the scattering parameters characterize the ability of particles to cross the shock and enter the acceleration process.},
and is determined stochastically for each particle in the \mc\ simulation.

A further assumption is that the subshock is transparent, i.e., the \mc\ model ignores any possible cross-shock potential or effects from enhanced magnetic turbulence, which would influence the subshock crossing probability or produce an energy change in the subshock layer. 
We emphasize that we do not present our injection model  as a ``solution" to the injection problem. Rather, it is a clearly defined process that requires no additional assumptions beyond the scattering model basic to the \mc\ simulation. Within statistical uncertainties, it produces unique and \SC\ shock solutions that can be compared with observations \citep[e.g.,][]{EMP90,BOEF97}
or PIC simulation results \citep[e.g.,][]{EGBS93}. 
Of course, the downstream return probability and the injection fraction will change substantially if an oblique magnetic field and compressed turbulence
is taken into account 
\citep[e.g.,][]{NiemiecEtal2006,LR2006}. In principle, the 
smooth-shock techniques we demonstrate here can be applied to this far more complicated situation.

In Figure~\ref{fig:Inj_frac} we show how the injection fraction varies with $\Ng$ for two UM shocks, one with $\gamZ=1.5$ (top panel) and one with $\gamZ=10$ (bottom panel). 
The choice of $N_g=15$ implies $\delMax \simeq \pi/2$ 
(i.e., equation~\ref{eq:Tmax}) and is near the extreme LAS limit. 
Larger values of $N_g$ imply fine PAS. 
The fraction of protons that have crossed the shock three times (up to down, down to up, and up to down again) is clearly indicated by the sharp change in the spectral shape at $p/(m_p c) \sim \gamZ$. 

For the \transrel\ shock ($\gamZ=1.5$) there is little effect from $\Ng$, as expected, since $\Ng$ is known to be unimportant for \nonrel, parallel shocks. For $\gamZ=10$, $\Ng$ does influence the injection fraction and the acceleration. 
For the $\gamZ=10$ cases in the bottom panel of 
Figure~\ref{fig:Inj_frac}, 
the smallest value of $N_g$ results in 
fewer particles injected after the first shock crossing.
However the flatness of the $N_g=15$ and $34$ curves above $p\sim 10 m_p c$ show that more particles continue to be accelerated for low $N_g$ compared to higher $N_g$. 
With a small $N_g$, particles move a larger distance between scattering events and will first interact with the downstream plasma further from the discontinuous shock (or subshock as the case may be) than with fine PAS. This reduces the probability that these particles will return upstream. Countering this effect, however, is the fact that when $\gamZ$ is large, LAS produces larger energy gains, on average, than fine PAS for subsequent shock crossings. This can produce a harder spectrum, as seen in Figure~\ref{fig:Inj_frac}
\citep[see][ for a detailed discussion of this effect]{ED2004}. 
While the above effect is present in the $\gamZ=1.5$ plots, it is much less noticeable. 

What is not shown in the UM examples in 
Figure~\ref{fig:Inj_frac} is that, 
as the shock becomes modified by the backreaction from accelerated particles,
the injection fraction at the subshock automatically adjusts 
according to the change in the subshock compression and the 
pre-heating that occurs in the shock precursor.

\section{Results} \label{sec:Results}
We now show detailed results for shock speeds spanning the range from \nonrel\ to \ultrarel, i.e., 
(A) $\betaZ=0.0667$ (i.e., $u_0=2\xx{4}$\,\kmps);
(B) $\betaZ = 0.2$;
(C) $\gamZ=1.5$;
(D) $\gamZ=10$; and
(E) $\gamZ=30$.

\subsection{Unmodified Shocks}
\begin{figure}
\epsscale{1.05}
\plotone{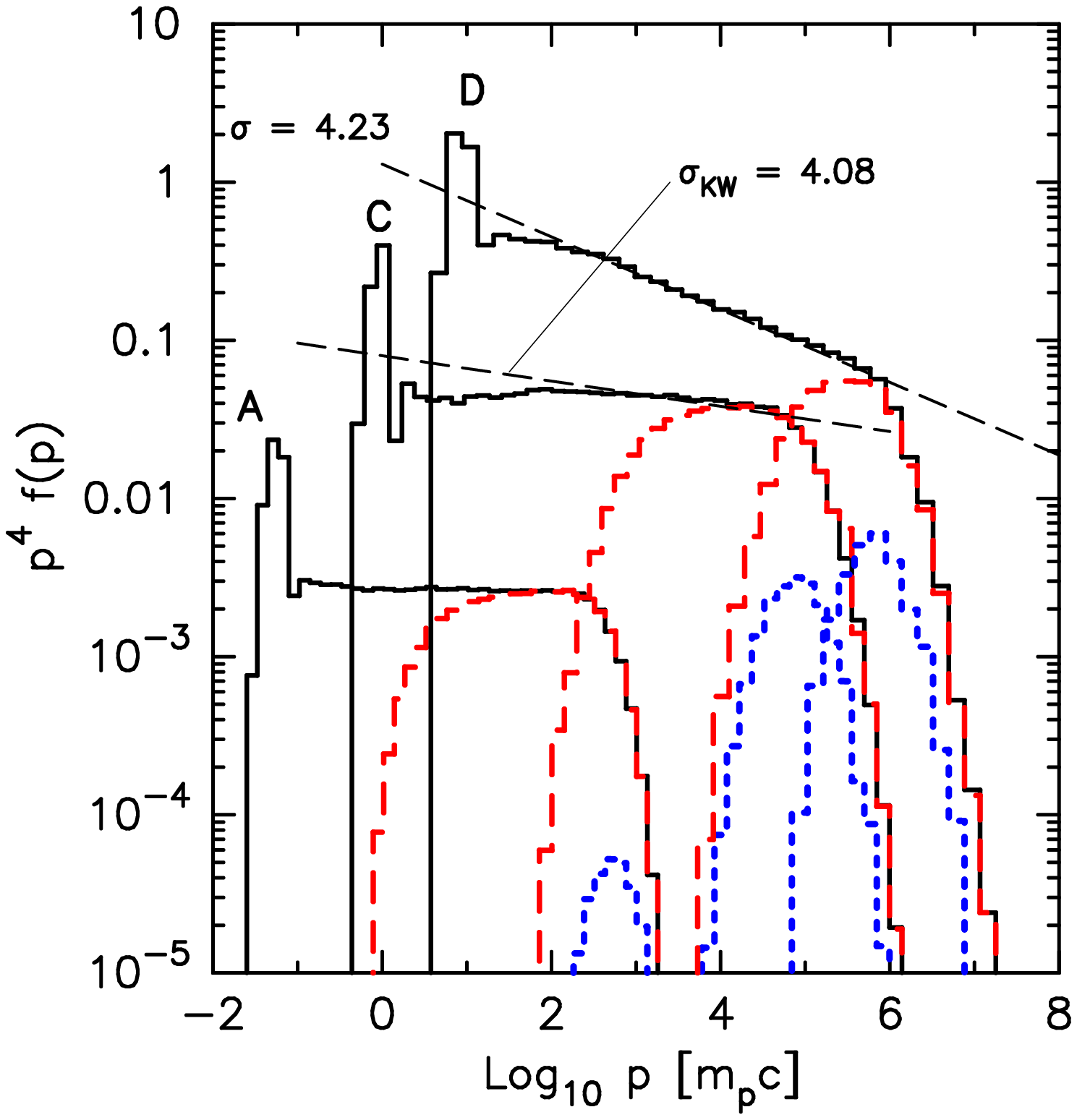} 
\caption{Omni-directional spectra measured in the shock frame
from three \Unmod\ shocks: $u_0=2\xx{4}$\,\kmps (A), $\gamZ=1.5$ (C), and $\gamZ=10$ (D).  
All spectra are scaled to an injected flux of 1 proton-cm$^{-2}$-s$^{-1}$ and, for all models, 
the solid (black) spectra were calculated downstream from the shock, the dashed (red) spectra were calculated at $x=-100\rgZ$, and the dotted (blue) spectra were calculated at the FEB. 
The positions of the FEB for the three shocks are: $-3\xx{4}\rgZ$ (A), $-10^4\rgZ$ (C), and $-10^3\rgZ$ (D).
The thermal portions of the upstream spectra are not shown for clarity.
The dashed line labeled $\sigKW$ is the spectral index determined from \citet{KW2005}.
Note that the vertical axis is $p^{4} f(p)$.
\label{fig:UM_nonrel_rel}}
\end{figure}

In Fig.~\ref{fig:UM_nonrel_rel} we show spectra from UM shocks for \nonrel\ ($u_0=2\xx{4}$\,\kmps; Model A), \transrel\ ($\gamZ =1.5$; Model C), and fully \rel\ ($\gamZ=10$; Model D) shock speeds. Other parameters for these shocks are given in Table~\ref{tab:Input}. 
For the examples in Fig.~\ref{fig:UM_nonrel_rel}  we have chosen $\Ng$ large enough to obtain convergent solutions and, for each case, we plot spectra calculated at different positions relative to the 
subshock at $x=0$.\footnote{By convergent we mean that no changes other than statistical fluctuations occur if $N_g$ is increased further.} 
The  solid (black) curves  are  calculated downstream from the subshock, the dashed (red) curves are calculated upstream from the subshock at $x=-100\,\rgz$, and the dotted (blue) curves were calculated at the upstream FEB.
The gyroradius $\rgz \equiv m_p u_0 c /(e B_0)$ varies with shock speed. 
These spectra, and all other $f(p)$ plots,  are shock-frame, 
omni-directional, phase-space distributions.\footnote{A shock-frame, 
omni-directional spectrum is one that would be measured by a 
$4\pi$-steradian detector stationary in the shock frame.}

Now consider the spectra calculated downstream from the shock (black curves in Fig.~\ref{fig:UM_nonrel_rel}). 
For the \nonrel\ shock, the particle distribution function between the thermal peak and the turnover produced by the FEB is well matched by a power law with spectral index $\sigma \simeq 4$ (i.e., $f(p)d^3p \propto p^{-\sigma} d^3p$), as expected.
Also as expected, the fully \rel\ case with $\gamZ=10$ obtains a power law in the mid-energy range with $\sigma \simeq 4.23$, the well known \ultrarel\ result \citep[e.g.,][]{BO98,KirkEtal2000}. For the \transrel\ case 
($\gamZ=1.5$, $\Rtot \simeq 3.53$, Model C), we compare the \MC\ result against the expression from 
\citet{KW2005}, i.e., $\sigKW\simeq 4.08$ (see their equation 23) and find reasonable agreement for the power-law index in the 
range $p \gtrsim 300\,m_p c$ and below the turnover from the FEB. 

The dashed and dot-dashed
(red and blue) spectra calculated in the shock precursor are missing the lowest energy particles since these are not able to diffuse back upstream to the observation position.\footnote{The thermal injected particles that have not yet crossed the subshock are present at upstream positions but are not shown in
Figure~\ref{fig:UM_nonrel_rel} for clarity.}
The particles that escape at the FEB (dotted, blue curves) 
are streaming freely away from the shock which causes the 
omni-directional distributions to be depressed relative to the more isotropic fluxes well within the precursor. 
We discuss this point with the \SC\ shock solutions  shown next.

\subsection{Nonlinear Relativistic Shocks}
If astrophysical shocks are efficient particle accelerators the NL backreaction of the accelerated particles on the shock structure must be taken into account \citep[e.g.,][]{Drury83,BE87}. To our knowledge, apart from PIC simulations, the only previous attempt to treat NL, \rel\ shocks was the preliminary work of \citet{ED2002}.

\subsubsection{Fine scattering solutions} \label{sec:fine_scat}
We now show \SC, fine scattering  shock solutions.
Again, a ``fine" scattering solution has $N_g$ large  enough so no changes, other than statistical, result if $N_g$ is increased.

\begin{figure}
\epsscale{1.05}
\plotone{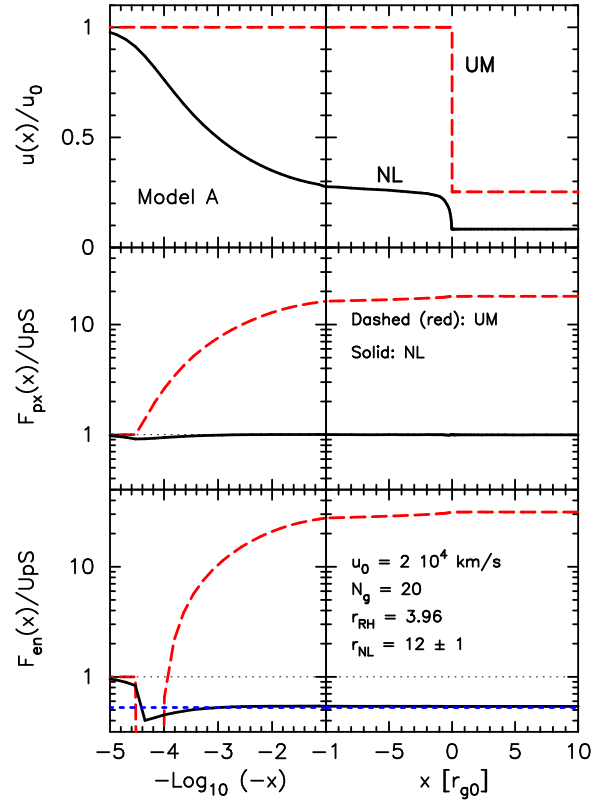} 
\caption{The top panels show the shock structure in terms of $u(x)/u_0$ (solid-black and dashed-red curves)  vs. position $x$ relative to the subshock at $x=0$. 
Note the split log--linear scale.
Length is measured in units of 
$\rgZ= m_p u_0 c /(e B_0)$. 
The middle panels show the momentum flux $\Fpx$ and the bottom panels 
show the energy flux $\Fen$. Values in all panels are  
scaled to far upstream values. 
In all panels, the dashed (red) curves are from an UM shock 
with the same input values as the modified NL shock results 
shown with solid (black) curves. 
For this \nonrel\ shock, the \SC\ compression ratio 
is $\rNL = 12\pm 1$ compared to the \RH\ value $\rRH \simeq 3.96$.
The thick horizontal dotted (blue) line in the bottom panels at $\sim 0.5$
shows the energy flux expected from our consistency conditions derived from equations~(\ref{eq:NumFlux})--(\ref{eq:EnFlux}). The FEB is  at $x = - 3\xx{4}\,\rgZ$.
\label{fig:profile_Vsk2e04}}
\end{figure}

Figure~\ref{fig:profile_Vsk2e04} shows the shock structure for $u_0=2\xx{4}$\,\kmps\ (Model A). The dashed (red) curves show the UM solution with $\rRH \simeq 3.96$ and it is clear from the lower panels that the momentum and energy fluxes are not conserved by large factors. 
Of course,  the amount of acceleration depends on the 
injection model used, so these results are particular to the thermal leakage model contained in the \mc\ simulation.
The solid (black)
curves show the \SC\ result obtained by iteration
where the shock has been smoothed and the compression ratio has been increased to $\rNL = 12 \pm 1$. Now, once the escaping energy flux at the FEB is accounted for, the momentum and energy fluxes are conserved to within a few percent of the far upstream values. The heavy
dotted (blue) horizontal line in the energy flux  panels is the fractional energy flux 
(i.e., $1 - \QescEn/\FenZ$), calculated from 
equations~(\ref{eq:NumFlux})--(\ref{eq:EnFlux})
once  the modification of the downstream ratio of specific heats has been accounted for. Except for a small deviation near the 
FEB at $x=-3\xx{4}\,\rgZ$, 
the \mc\ result is consistent with 
equations~(\ref{eq:NumFlux})--(\ref{eq:EnFlux}). The shock structure shown in the top panels, with $\rNL = 12 \pm 1$, is a consistent solution where approximately 50\% of the energy flux escapes at the upstream FEB. There is essentially no escaping momentum flux as expected for a \nonrel\ shock \citep[see][]{Ellison85}.

Solutions with $\rNL \gg \rRH$, such as shown in 
Figure~\ref{fig:profile_Vsk2e04}, have been known for some time  for \nonrel\ shocks \citep[e.g.,][]{JE91,BE99} and our $u_0 = 2\xx{4}$\,\kmps\ result is consistent with these previous results.

\begin{figure}
\epsscale{1.0}
\plotone{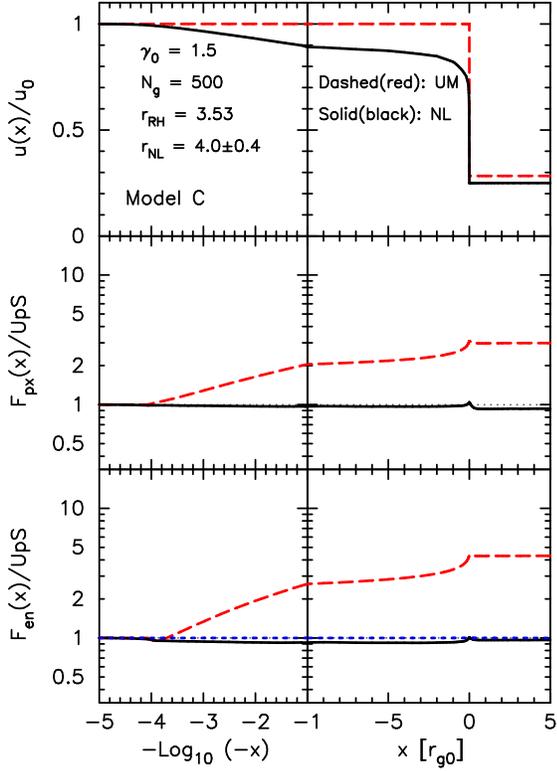} 
\caption{Same format as Figure~\ref{fig:profile_Vsk2e04} for $\gamZ=1.5$ (Model C). In all panels the dashed (red) curves show the UM case and the solid (black) curves show the \SC\ case. For this example, the expected escaping momentum and energy  fluxes predicted from our consistency conditions are small, as indicated by the horizontal dotted (blue) line in the bottom panels.
\label{fig:profile_gam1.5}}
\end{figure}

In Figure~\ref{fig:profile_gam1.5} we show the UM and NL shock structures for our example (C) with $\gamZ=1.5$. The NL effects are present but less dramatic than for $u_0=2\xx{4}$\,\kmps. For $\gamZ=1.5$, the \SC\ compression ratio is $\rNL=4.0 \pm 0.4$ compared to $\rRH\simeq 3.53$. 
There is an insignificant amount of escaping  flux through the FEB at $x=-10^4\,\rgZ$
and the momentum and energy fluxes across the shock are equal to the far upstream values to within a few percent.

For this case, the increase in $\Rtot$ comes about because the downstream ratio of specific heats, $\GamTwo$,  
decreases from the \RH\ 
value when particles are accelerated. This decrease makes the shocked plasma more compressible and results in $\rNL > \rRH$. 
The downstream $\GamTwo$ decreases from $\GamTwo \simeq 1.51$, the value obtained from equation~(\ref{eq:SpHeat}), to 
$\GamTwoMC \simeq 1.47$ in the NL case, enough of a difference to explain the increased compression ratio.
For the parameters we are using, $\gamZ=1.5$ is a transitional speed below which strong NL effects become dominant in the fine-scattering limit.

\begin{figure}
\epsscale{1.0}
\plotone{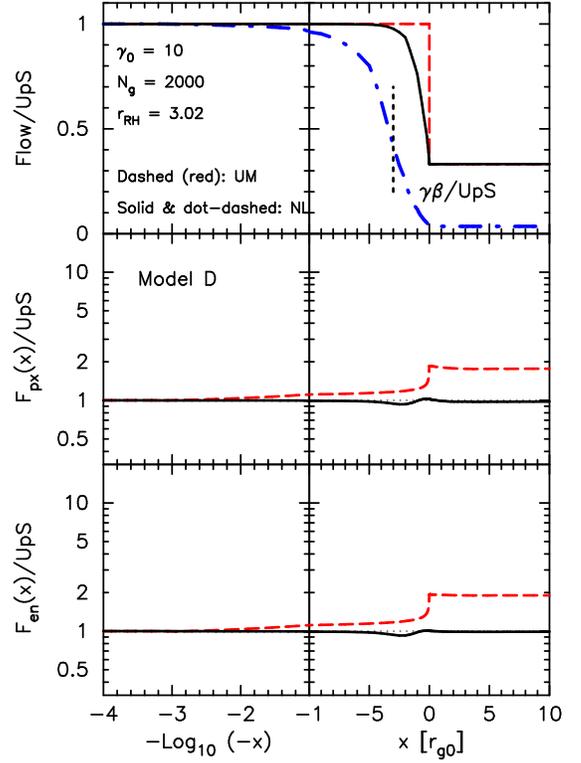} 
\caption{Same format as Figures~\ref{fig:profile_Vsk2e04} and 
\ref{fig:profile_gam1.5} for $\gamZ=10$ (Model D).
In all panels the dashed (red) curves show the UM case and the solid (black) curves show the \SC\ case.
To emphasize the shock smoothing, we show $\gamma(x) \beta(x)/(\gamZ \betaZ)$ in the top panels with 
a dot-dashed (blue) curve. The vertical dotted line in the top right panel shows the position of $\Xsub$.
\label{fig:profile_gam10}}
\end{figure}

In Figure~\ref{fig:profile_gam10} we show the shock structures for our fully \rel\ example (D) with $\gamZ=10$. For this case, $\rNL \simeq \rRH \simeq 3.02$. There are no significant escaping energy or momentum fluxes at the FEB and,  since the shocked particles are fully \rel\ without additional acceleration, $\GamTwo \simeq 4/3$. Thus, in this fine-scattering example, there are no mechanisms to increase $\Rtot$ above $\rRH$.
Nevertheless,
the energy and momentum fluxes in the UM shock are nearly a factor of two above the far upstream values.
A noticeable smoothing of the shock is necessary to conserve momentum and energy as shown by 
the $\gamma(x) \beta(x)/(\gamZ \betaZ)$ curve (dot-dashed, blue) in the top panels of
Figure~\ref{fig:profile_gam10}.

The smoothing parameter $\Xsub=-3\,\rgz$ is shown as a vertical dotted line in the top right panel of Figure~\ref{fig:profile_gam10}. 
The flow profile to the right of this line is determined by equation~(\ref{eq:Xsub}).
This is matched to the profile to the left by our algorithm based on 
equation~(\ref{eq:Pmc}).
The procedure is iterated until consistent momentum and energy fluxes are obtained as shown by the solid (black) curves in the lower panels. 

\begin{figure}
\epsscale{1.0}
\plotone{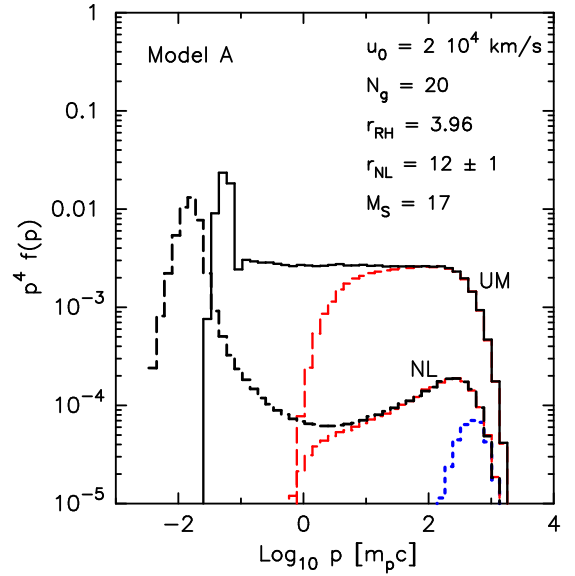} 
\caption{Omni-directional, shock frame spectra for 
$u_0=2\xx{4}$\,\kmps\ (Model A). 
The curves with the thermal peaks are  calculated behind the shock and the dashed (red) curves are calculated in the shock precursor at $x=-100\rgZ$. The dotted (blue) curve is the NL spectrum calculated at the FEB. The UM spectra are the same as shown in 
Figure~\ref{fig:UM_nonrel_rel}.
\label{fig:fp_Vsk2e04}}
\end{figure}

In Figure~\ref{fig:fp_Vsk2e04} we compare the UM and NL phase-space
distribution functions  for our $u_0=2\xx{4}$\,\kmps\ models. There is a dramatic difference with the superthermal
NL spectra being less intense and showing the characteristic concave upward shape. The shift in the thermal peak to lower energy, which accompanies efficient DSA, is also clear in the \SC\ result. This shift is a robust prediction of DSA. When acceleration occurs,
the subshock must weaken and injection must decrease to conserve energy and momentum.

We note that our \nonrel\ example has a sonic Mach number of $M_S \sim 17$ which is unrealistically low for such a high-speed shock in the typical interstellar medium. We have chosen parameters to yield a low $M_S$ for computational convenience. As shown in \citet{BE99}, the \SC\ compression ratio is a fairly strong
function of $M_S$ (i.e., $\Rtot \sim 1.3 M_S^{3/4}$) when  only adiabatic heating occurs in the precursor, as we assume here. Such high $\Rtot$'s make finding a \SC\ solution difficult and are almost certainly unrealistic in any case, since \alf\ wave damping is likely to be a source of heating in the precursor. The sonic Mach number becomes much less important for \rel\ shocks unless the upstream temperature is high enough for the unshocked particles to have a Lorentz factor $\sim \gamZ$. We do not consider such a situation here.

The distribution of escaping particles for the NL \nonrel\ shock is shown as a dotted (blue) curve in Figure~\ref{fig:fp_Vsk2e04}. As shown in Figure~\ref{fig:profile_Vsk2e04}, this distribution contains $\sim 50\%$ of the total energy flux but this fraction cannot be directly obtained from the omni-directional distributions we plot. 
(In Figure~\ref{fig:eff_nonrel_to_rel} below we give a direct result for the acceleration efficiency.)
What 
Figure~\ref{fig:fp_Vsk2e04} does clearly show is that the shape of the escaping distribution is dramatically different from the downstream distribution.

\begin{figure}
\epsscale{1.0}
\plotone{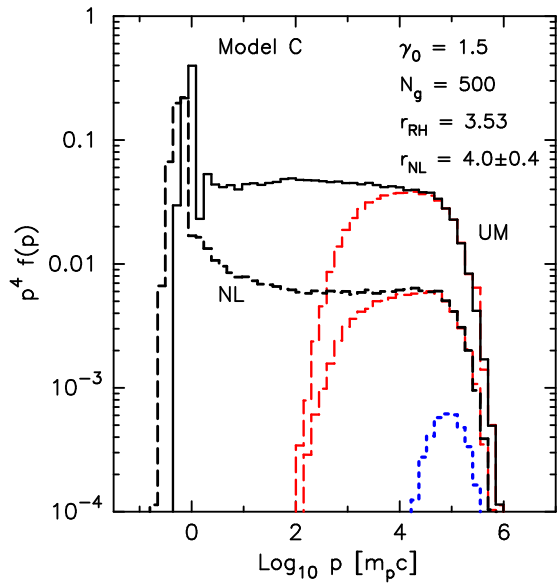} 
\caption{Comparison of UM shock spectra (Model C; same as in 
Figure~\ref{fig:UM_nonrel_rel}) with NL shock spectra. The shock structures for these $\gamZ=1.5$ shocks are shown in 
Figure~\ref{fig:profile_gam1.5}. The 
dashed (red) curves are calculated in the shock precursor at $x=-100\rgZ$. The dotted (blue) curve is the NL spectrum calculated at the FEB.
\label{fig:fp_gam1.5}}
\end{figure}

In Figure~\ref{fig:fp_gam1.5} we show the UM and NL distribution functions for our \transrel\ example with $\gamZ=1.5$. The differences between the UM and NL spectra are less than for $u_0=2\xx{4}$\,\kmps\ but they are still significant. The normalization of the superthermal NL distribution is approximately an order of magnitude less than the UM case and there is some slight concave curvature from the shock smoothing. 
While the thermal peak still shifts to a lower energy in the NL case, this shift is much less than it was for the NL \nonrel\ shock shown in Figure~\ref{fig:fp_Vsk2e04}.

The escaping flux is too small to influence the shock dynamics.
This is evident both from Figure~\ref{fig:profile_gam1.5} and from the distribution of escaping particles, shown in Figure~\ref{fig:fp_gam1.5} as a dotted (blue) curve. The normalization of the escaping particle distribution relative to the downstream distribution is more than an order of magnitude less than it is in the \nonrel\ case (c.f. Figure~\ref{fig:fp_Vsk2e04}).

\begin{figure}
\epsscale{1.0}
\plotone{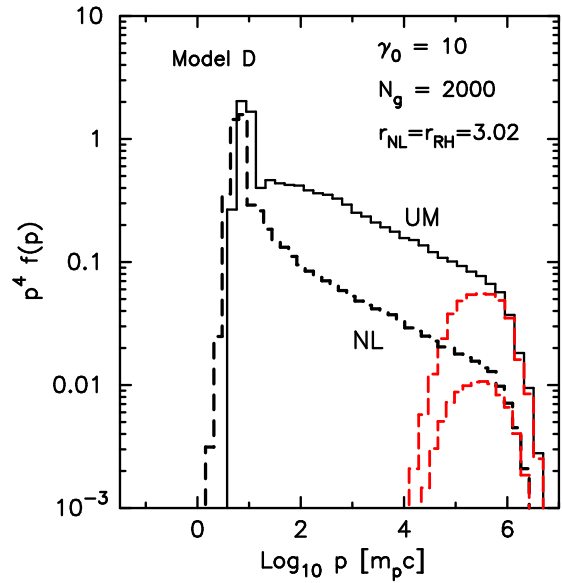} 
\caption{Comparison of UM shock spectrum (Model D; same as in 
Figure~\ref{fig:UM_nonrel_rel}) with NL shock spectrum. The shock structures for these $\gamZ=10$ shocks are shown in 
Figure~\ref{fig:profile_gam10}. The solid (black curves)  are calculated downstream from the subshock in the shock rest frame and the 
dashed (red) curves are calculated in the shock precursor at $x=-100\rgZ$.
\label{fig:fp_gam10}}
\end{figure}

The distribution functions for our $\gamZ=10$ example (Model D) are shown in Figure~\ref{fig:fp_gam10}.
While the shapes of the UM and NL spectra are quite similar, 
they both obtain $f(p) \sim p^{-4.23}$ power laws above the thermal peak,
the normalization drops by a factor of $\sim 5$ when energy and momentum conservation are taken into account.
There is also a slight shift in the thermal peak to lower energy as required when DSA occurs.

\begin{figure}
\epsscale{1.05}
\plotone{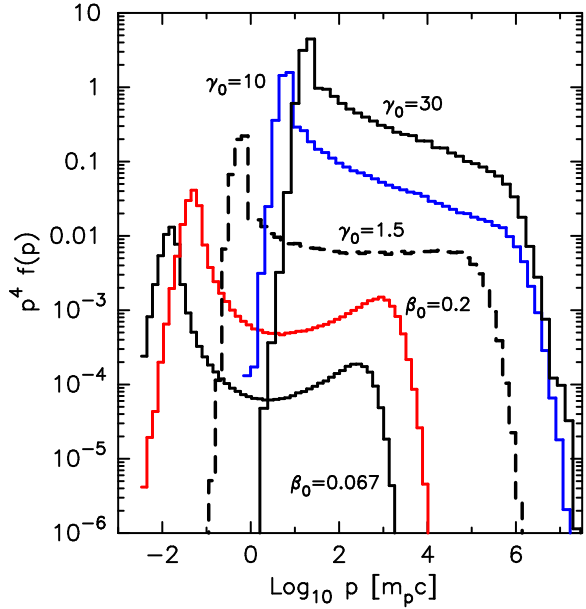} 
\caption{Nonlinear particle distributions calculated downstream from the shock in the shock rest frame for various shock speeds as indicated (Models A--E in Table~\ref{tab:Input}). The spectrum for the $\gamZ=1.5$ shock (dashed black curve) shows the transitional nature of \NL\ DSA. For speeds faster than $\gamZ \sim 1.5$, NL effects are relatively minor. For speeds below $\gamZ \sim 1.5$, strong NL effects occur and concave spectra are produced.
The strong increase in normalization from $\betaZ=0.067$ to $\gamZ=30$ mainly results from the fact that thermal particles gain more energy in their first shock crossing as the shock speed increases. The maximum momentum is determined by the position of the FEB in the various shocks.
\label{fig:fp_nonrel_to_rel}}
\end{figure}

In Figure~\ref{fig:fp_nonrel_to_rel} we compare the \SC\ distribution functions for our five examples spanning the range from $\betaZ=0.067$ to $\gamZ=30$.  The transitional character of the spectrum at $\gamZ=1.5$ is clear. At shock speeds below $\gamZ \sim 1.5$, DSA can be extremely efficient and strong \NL\ effects from particle escape and $\rNL > \rRH$ result in concave upward
spectra which are harder than $p^{-4}$ above a few GeV. 
These slower shocks will be strongly dependent on the input conditions, i.e., Mach number, the injection model, the position of the FEB which determines $\pmax$, assumptions concerning magnetic turbulence production and damping, and a finite speed for the magnetic scattering centers
\citep[e.g.,][]{BE99,VBE2008}. 
Above $\gamZ \sim 1.5$, the modified shocks have compression ratios $\sim \rRH$ and produce spectra that are nearly $\sigma = 4.23$ power laws. These faster shocks will be strongly dependent on the details of particle scattering (our parameter $\Ng$ discussed below), and on effects from oblique geometry and compressed downstream magnetic fields which we don't model.

\begin{figure}
\epsscale{1.0}
\plotone{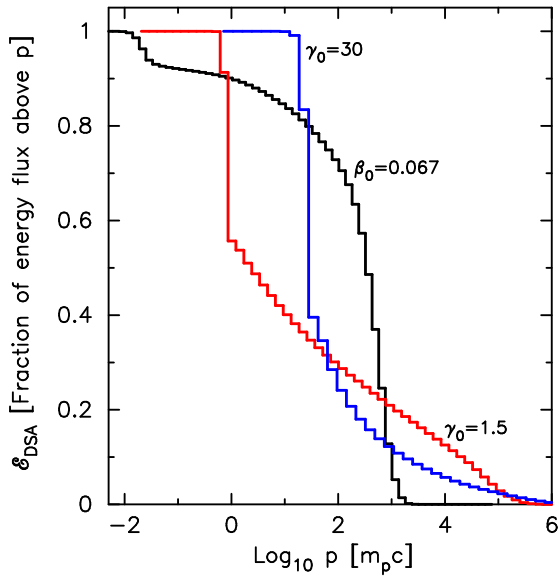} 
\caption{Acceleration efficiency  
in terms of the fraction of energy flux above shock frame
$p$ for the $\betaZ=0.067$ (A), $\gamZ=1.5$ (C), and $\gamZ=30$ (E) shocks shown in Figure~\ref{fig:fp_nonrel_to_rel}. These fractions include the energy flux that escapes at the upstream FEB. The escaping fraction is $\sim 0.5$ for $\betaZ=0.067$ but $\lsim 0.02$ for the two other cases. 
\label{fig:eff_nonrel_to_rel}}
\end{figure}

\subsubsection{Acceleration efficiency}
We define the acceleration efficiency, $\EffDSA(>p)$, as the fraction of total energy flux placed in particles with shock-frame momentum $p$ or greater. This is shown in 
Figure~\ref{fig:eff_nonrel_to_rel} for  three of the spectra shown in 
Figure~\ref{fig:fp_nonrel_to_rel}. The other two lie between these examples. These fractions include the escaping energy flux.

If the fraction of energy above the downstream thermal peak is the measure of efficiency, all of these shocks are extremely efficient. 
The \nonrel\ 
shock ($\betaZ=0.067$, black curve) places over 90\% of the incoming energy flux into superthermal particles. Approximately 50\% of the total flux is in escaping particles.
While the more \rel\ shocks are less efficient, the energy flux in particles with Lorentz factors  $\gsim \gamZ$, is still large. $\EffDSA \sim 0.55$ for $\gamZ=1.5$ and
$\EffDSA \sim 0.35$ for $\gamZ=30$.
For these cases, the escaping energy flux is insignificant.
Additional factors, such as magnetic turbulence generation and wave damping, are not included here and will reduce
the acceleration efficiency
\citep[e.g.,][]{BE99,CBAV2009}.

\subsubsection{Large-angle-scattering solutions}
\begin{figure}
\epsscale{1.0}
\plotone{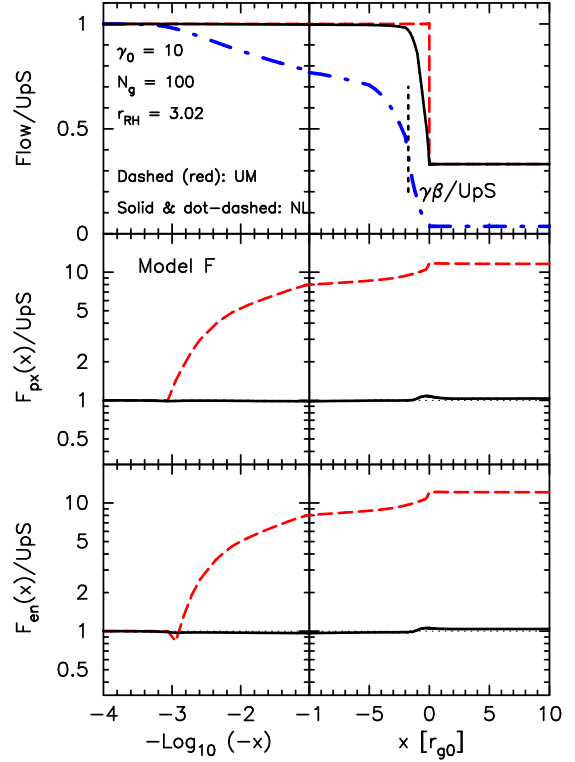} 
\caption{Shock structure for \LAS\ Model F with $\gamZ=10$ and $N_g=100$. All input parameters for this model are the same as Model D (Figure~\ref{fig:profile_gam10}) except $N_g$.
The figure format is the same as in 
Figure~\ref{fig:profile_gam10}.
\label{fig:profile_Gam10_Ng100}}
\end{figure}

The fineness of scattering, as determined by our parameter $\Ng$, strongly influences DSA in \rel\ shocks. This effect is discussed in \citet{ED2004} where results for UM shocks with 
large-angle-scattering (LAS) (i.e., $\Ng$ much less than the value needed for convergence) are shown. 
Large-angle-scattering produces hard spectra and implies efficient DSA so NL effects are certain to be important. 
Figure~\ref{fig:profile_Gam10_Ng100} (Model F) 
shows the dramatic effect shock smoothing has on a $\gamZ=10$ shock with $\Ng =100$ (versus $\Ng =2000$ used to obtain the convergent $\sigma \simeq 4.23$ solution shown in Figure~\ref{fig:profile_gam10}). The only difference in input parameters between this case and the shock shown in 
Figure~\ref{fig:profile_gam10} is $\Ng$.

\begin{figure}
\epsscale{1.0}
\plotone{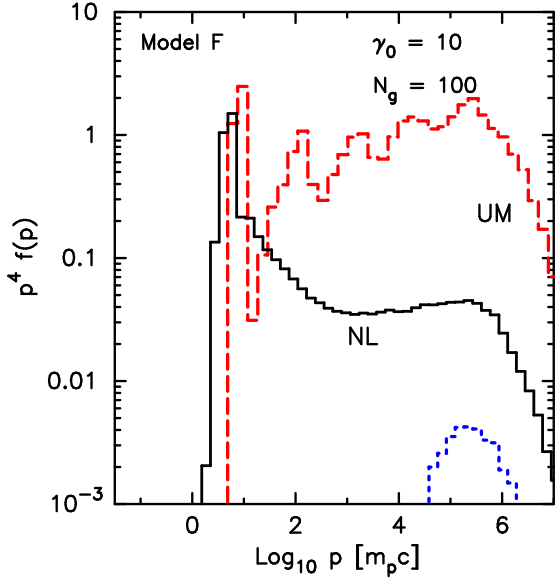} 
\caption{Omni-directional, shock frame spectra for a $\gamZ=10$ shock with \LAS, i.e., $\Ng=100$. The solid (black) and dashed (red) curves are measured downstream from the shock and the dotted (blue) curve is the spectrum measured at the FEB for the NL shock. The peaks in the dashed (red) curve reflect discrete shock crossings in the UM shock. This coherence is removed in the smoothed NL shock.
\label{fig:fp_gam10_Ng100}}
\end{figure}

With LAS the energy flux is out of conservation by a factor of $\sim 10$ with no shock smoothing and the UM and NL particle distribution functions are vastly different as shown in  
the downstream spectra plotted in Figure~\ref{fig:fp_gam10_Ng100}.
In the UM case (dashed, red curve), discrete peaks are present in $f(p)$ corresponding to a given number of shock crossings. Similar spectra were seen in 
\citet*{EJR90} and recently in \citet{SummerlinBaring2012}.
The lowest momentum peak at $p/(m_pc) \simeq \gamZ$ is from  particles that have crossed the shock once. In addition,  clear peaks for three, five, and seven crossings are present, each at a momentum approximately a Lorentz factor greater then the previous peak.

The  consistent NL solution is shown as the solid (black) curve in 
Figure~\ref{fig:fp_gam10_Ng100}. 
The peaks from discrete shock crossings  seen in the UM shock are gone.
The superthermal part of the spectrum is much softer, the lowest momentum peak is at a slightly lower momentum than in the UM case, and the rest of $p^4 f(p)$ shows the  concave upward curvature which is seen in  efficient DSA in \nonrel\ shocks.
This concave curvature is present in Fig.~\ref{fig:fp_gam10} but it is less evident because the acceleration is less efficient with $\Ng =2000$.
In Figure~\ref{fig:fp_gam10_Ng100}, the NL spectrum obtains a
power-law index just below the turnover caused by the FEB which is slightly harder than $p^{-4}$.
The NL result conserves energy and momentum and this sets the normalization of $f(p)$ which is almost a factor of 100 below the UM $f(p)$ at $p=10^6\,m_p c$.

\section{Discussion} \label{sec:Diss}
From kinematics, \ultrarel\ shocks are certain to energize incoming cold particles to 
$p \sim \gamZ m_0 c$ in the shock frame
(see the ``thermal peak" in any of our  spectral 
plots). Diffusive
shock acceleration beyond this, where particles are required to make repeated shock crossings, is much less certain. 

\begin{figure}
\epsscale{1.0}
\plotone{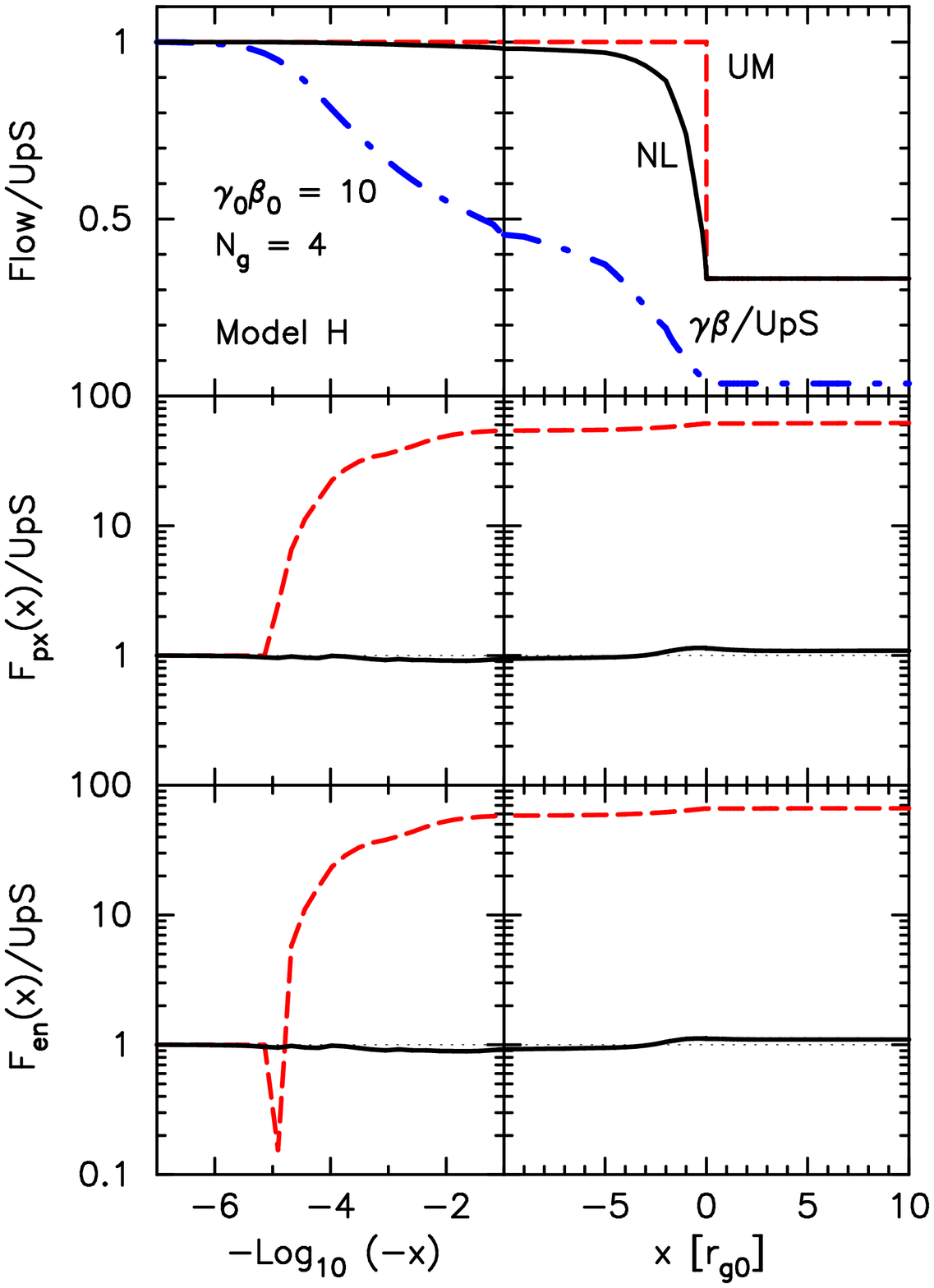} 
\caption{The top panel shows the shock structure for Model H matching the \citet{SummerlinBaring2012} example discussed in the text.
The dashed (red) curve is the UM flow speed $u(x)$, the solid (black) curve is the \SC\ flow speed, and the dot-dashed (blue) curve is $\gamma(x)\beta(x)/(\gamZ \betaZ)$ for the NL shock. In the middle and bottom panels, the dashed (red) curve is the UM flux, 
and the solid (black) curve is the NL flux. The momentum and energy fluxes are conserved to within a few percent in the \SC\ shock.
\label{fig:Grid_Baring_Ng4}}
\end{figure}

\begin{figure}
\epsscale{1.0}
\plotone{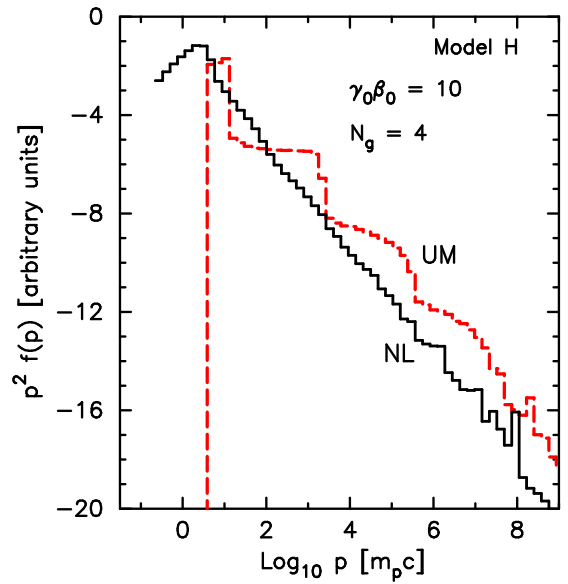} 
\caption{The dashed (red) curve is the downstream spectrum from the UM shock shown in Figure~\ref{fig:Grid_Baring_Ng4}. The solid (black) curve is the NL spectrum. Note that we plot $p^2 f(p)$ to match 
Figure~12 of \citet{SummerlinBaring2012} and the normalization of the vertical-axis is arbitrary.
\label{fig:fp_Baring_Ng4}}
\end{figure}

\subsection{Comparison with Previous \MC\ Results}
\MC\ techniques have been used to study \rel\ shocks for some time \citep[e.g.,][]{Ostrowski1988,Ostrowski1993,EJR90,BO98,Baring1999}.
Recently, using essentially the same scattering assumptions employed here, \citet{SummerlinBaring2012} made a comprehensive study of the plasma parameters important for
\TP\ DSA, including the shock obliquity.
Here we compare our NL results (Model H)
with the $\Tbn=0^\circ$, $\gamZ \betaZ = 10$, $\Tscat=\delMax = \pi$ example shown in Figure~12 of 
\citet{SummerlinBaring2012} and in 
Figure~2 of \citet{SteckerEtal2007}.
This LAS example, where $\delMax=\pi$ implies $N_g\simeq 4$ 
from equation~(\ref{eq:delMax}), is similar to, but more extreme than, our Model F shown in Figures~\ref{fig:profile_Gam10_Ng100} and 
\ref{fig:fp_gam10_Ng100}.

A qualitative comparison with the \TP\ \citet{SummerlinBaring2012} result is sufficient to show the important differences that occur when a \SC\ shock structure is 
considered.\footnote{Superficial differences between our Model H and 
the $\Tscat = \pi$ model of \citet{SummerlinBaring2012} include the facts that we use a FEB to limit the acceleration rather than a maximum momentum, and that our upstream proton temperature ($T_0= 10^6$\,K) is less, as indicated by the narrow thermal peak in our UM spectrum versus the broader peak in the \citet{SummerlinBaring2012} spectrum. Neither of these differences influences our comparison in any important way.}
In Figure~\ref{fig:Grid_Baring_Ng4} we show the shock structure and momentum and energy fluxes with (solid black curves) and without 
(dashed red curves) considering the backreaction of accelerated particles. The extreme LAS assumption produces very efficient acceleration in the UM shock and the momentum and energy fluxes are over-produced by nearly a factor of 100. 
Shock smoothing must occur unless the injection is reduced by other factors
to the point where essentially only the downstream thermal distribution is present. The dot-dashed (blue) curve shows the extent of smoothing required to give a \SC\ solution.

The UM particle distribution shown in Figure~\ref{fig:fp_Baring_Ng4} as a dashed (red) curve is similar to that shown in Figure~12 of \citet{SummerlinBaring2012}. It shows the characteristic step-like structure of LAS in UM shocks.
As shown by the solid black curve in Figure~\ref{fig:fp_Baring_Ng4}, this step-like structure disappears in the \SC\ shock. It resembles the steeper $\delMax = \lsim 20^\circ$ examples shown in Figure~12 of \citet{SummerlinBaring2012}.
Even if turbulence is present that can result in LAS in \rel\ shocks, basic conservation considerations preclude the formation of flat, step-like spectra.

In contrast to the simple scattering assumptions made here and 
in \citet{SummerlinBaring2012}, the \MC\ studies of 
\citet*{NiemiecEtal2006}, and references therein, start with a background magnetic field structure including magnetic field perturbations with various wave-power spectra in \ultrarel, 
oblique shocks. 
Instead of assuming \PAS, they calculate particle trajectories directly as particles move through the background field.
These \TP\ results are designed to see how the background field influences the spectral shape of the accelerated population; they  do not include shock smoothing or any feedback between particles and waves.

The main conclusion \citet*{NiemiecEtal2006} reach is that, for a wide variation of magnetic field structures, spectra are softer than the ``universal" $f(p) \propto p^{-4.2}$ power law and show cutoffs at energies below the maximum resonance energy determined by the background turbulence. They call into question the ability of DSA to produce the radiating electrons seen in superluminal
\rel\ shock sources such as extra-galactic radio jets and GRB afterglows, or for \ultrarel\ DSA to be the mechanism to produce UHECRs 
\citep[see][ who reach somewhat different conclusions, 
for similar \MC\ work on perpendicular shocks]{LR2006}.

Superluminal shocks require strong cross-field diffusion for DSA to occur.
In an oblique shock, 
the \HT\ (H-T) frame is the frame
where the $\mathrm{\bf u} \times \mathrm{\bf B}$ electric field is zero. This frame moves along the shock front at a speed 
$\vHT = \gamZ u_0 \tan{\Tbn}$ and shocks with $\vHT >c$ are superluminal. 
A downstream particle tied to a magnetic field line cannot recross the shock and be diffusively accelerated when $\vHT >c$.
\Ultrarel\ shocks will be superluminal unless 
$\Tbn < 1/\gamZ$.
For $\gamZ=100$, 
$\Tht \equiv \tan^{-1}{[1/(\betaZ\gamZ)]} \simeq 0.6\degg$, essentially 
 precluding DSA in plasmas where the turbulence is too weak to produce strong cross-field diffusion.
For $\gamZ=1.5$, $\Tht \simeq 42\degg$ and far more phase space is open to DSA even without cross-field diffusion. 
As we have emphasized above, all of the factors that limit the efficiency of DSA in \ultrarel\ shocks, including the wave structures investigated, for example, by \citet*{NiemiecEtal2006}, become less important as the shock slows to  \transrel\ speeds.

\subsection{Comparison with PIC Results}
The most compelling reason to believe that \ultrarel\ shocks do, in fact, accelerate particles beyond the initial kinematic
boost comes from PIC simulations 
\citep[e.g.,][]{Hoshino92,Kato2007,
SironiSpit2009,KeshetEtal2009,SironiSpit2011}.
These simulations clearly show that acceleration occurs under some circumstances 
with characteristics similar to that expected from DSA.
Keeping in mind the fundamental differences in how the  
wave-particle interactions and acceleration are treated in a PIC simulation versus a \mc\ simulation, we qualitatively compare our results to the $\gamZ=15$, $\theta=15\degg$ example shown in 
\citet{SironiSpit2011}. 
Here, $\theta$ is the angle between the far upstream magnetic field and the shock normal in the 
simulation (i.e., wall) frame of the PIC simulation. This obliquity 
is subluminal for $\gamZ=15$, making a comparison to our parallel shock results meaningful.

\begin{figure}
\epsscale{1.0}
\plotone{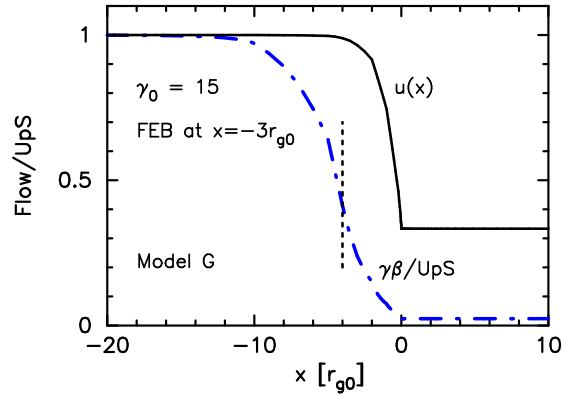} 
\caption{Nonlinear structure of a $\gamZ=15$ shock with a FEB at $x = -3\,\rgZ$ which conserves mass, momentum, and energy flux
(Model F). The vertical dotted line shows the position of $\Xsub=-4\,\rgZ$.
\label{fig:grid_gam15_FEB3}}
\end{figure}

In Figure~\ref{fig:grid_gam15_FEB3} we show \mc\ results for the structure of a NL shock with parameters chosen to produce spectra that can be compared to those in Figure~2 or 7 in \citet{SironiSpit2011}.
In order to obtain spectra with the low cutoff energy obtained 
in \citet{SironiSpit2011}, a short shock precursor is required.
We set the FEB at $x= -3\,\rgZ$ and the dot-dashed $\gamma \beta / (\gamZ \betaZ)$ curve shows that escaping particles influence the incoming flow upstream of $x= -3\,\rgZ$. The vertical dotted line shows the position of $\Xsub= -4\,\rgZ$ 
(i.e., equation~\ref{eq:Xsub}). For comparison, the FEB is at $x = -1000\,\rgZ$ for our $\gamZ=10$ Model D 
(Figures~\ref{fig:profile_gam10} and \ref{fig:fp_gam10}).

\begin{figure}
\epsscale{1.0}
\plotone{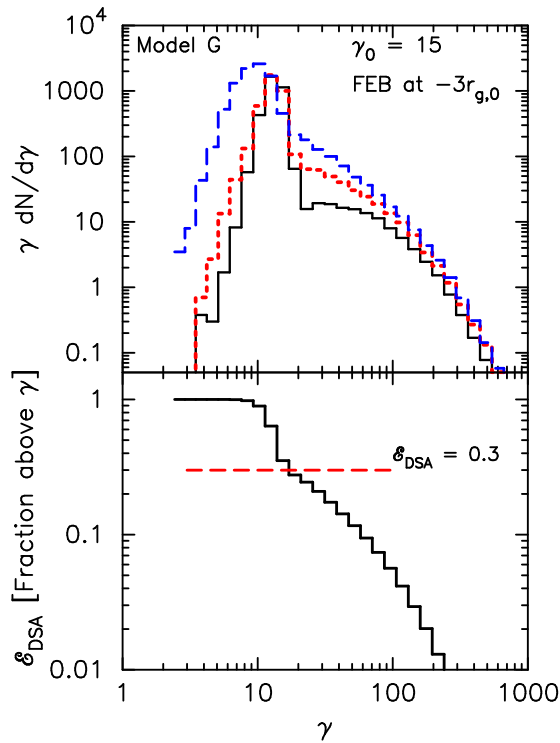} 
\caption{The top panel shows the  omni-directional, shock frame spectra for a NL $\gamZ=15$ shock with a FEB at $x = -3\,\rgZ$ 
(Model G). The solid (black) curve is measured at $x = -2.5\,\rgZ$, the dotted (red) curve is measured at $x = -2\,\rgZ$, 
and the dashed (blue) curve is measured downstream from the subshock at $x = 5\,\rgZ$. The bottom  panel shows the acceleration efficiency and the horizontal dashed line marks the efficiency found by 
\citet{SironiSpit2011} using a PIC simulation  for a shock with similar parameters.
\label{fig:fp_Acc_FEB3}}
\end{figure}

In the top panel of Figure~\ref{fig:fp_Acc_FEB3} we show \MC\ spectra measured downstream from the subshock (dashed, blue curve) and at two upstream positions; $x=-2\,\rgZ$ for the dotted (red) curve and $x=-2.5\,\rgZ$ for the solid (black) curve. Considering the different nature of the simulations 
we have not attempted to precisely match parameters for this illustration.
Nevertheless, the spectra in Figure~\ref{fig:fp_Acc_FEB3}, 
are in general
qualitative agreement with the proton spectra shown in Figure~2(g), (h), and (i) of 
\citet{SironiSpit2011}.\footnote{Note that our spectra are calculated in the shock frame and the \citet{SironiSpit2011} spectra are calculated in the  ``wall" or downstream frame. This does not make a large difference since the downstream flow Lorentz factor is $\sim 1$ in the shock frame.}  

\newlistroman

Common features between the PIC and \MC\ spectra are: 
\listromanDE  a broad thermal peak for the downstream spectrum with much sharper thermal peaks for the upstream spectra;
\listromanDE the downstream thermal peak is at $\gamma \simeq 10$ while the upstream peaks occur at $\gamma \simeq 15$ (if measured in the plasma frame, these upstream peaks would occur at $\sim \gamma^2$ here and in the PIC simulation);
\listromanDE a relatively smooth transition from thermal to superthermal energies in the downstream spectrum as thermal particles are injected into the acceleration mechanism;
\listromanDE a cutoff in the superthermal spectrum starting below $\gamma \sim 10^3$; and,
\listromanDE an acceleration efficiency (shown in 
the bottom panel of Figure~\ref{fig:fp_Acc_FEB3} for the \MC\ shock) of 
approximately 30\%.

\newlistroman

Important differences are:
\listromanDE that the downstream \MC\ spectrum is softer than the PIC one and doesn't flatten to  $dN/dg \propto \gamma^{-2.1}$ before the cutoff; and,
\listromanDE  that the spectral cutoff in the \MC\ result is broader than the PIC cutoff and occurs for a different reason
(the PIC  spectrum cuts off because of a finite run time, while the \MC\ spectrum cuts off because we have imposed a FEB at $x =-3\rgZ$ to produce a cutoff at an energy similar to that obtained in the PIC result).

Shock modification by the backpressure of accelerated particles
is also clearly present in both cases, as seen in our 
Figure~\ref{fig:grid_gam15_FEB3} and the top panel of
Figure~4 in \citet{SironiSpit2011}.\footnote{The density ratio in Figure~4 of \citet{SironiSpit2011} is $\sim 4$ because the densities in the plot are measured in the wall (i.e., downstream) rest
frame. When transformed to the shock frame, the density ratio is $\sim 3$ as expected.} 
Of course, the scaling of the shock structure, which depends on the magnetic field, is quite different. The \SC\ field in the PIC result is highly turbulent, time-dependent, and varies strongly with location. 
It also includes \MFA, 
presumably by Bell's instability 
\citep[e.g.,][]{Bell2004,Bell2005}.
In the \mc\ simulation Bohm diffusion is assumed and
the field is uniform 
and set to $B_0=3$\,\muG.
Nevertheless, shock modification must occur to satisfy energy and momentum conservation and the fact that the PIC and \MC\ shocks have similar acceleration efficiencies means the gross features of the shock structure will be similar.

Shock reformation is another notable aspect of the PIC results that is not modeled by our steady-state \mc\ simulation. In the 
time-dependent PIC simulations, bunches of downstream ions can return back upstream 
causing the subshock to disrupt and reform in a quasi-periodic fashion. 
While this process might be important for determining local particle reflection and other injection effects, the relatively narrow temporal and spatial averages applied by \citet{SironiSpit2011} in their Figures~4 and 7 for example seem to smooth the density profiles and energy spectra enough so long-term effects from reformation are  no longer evident.
This suggests that our steady-state \mc\ results can be 
meaningfully compared to averages taken during the evolving PIC simulations.

\subsection{GRB Afterglows}
The radiation observed from GRB afterglows is generally modeled as \syn\  emission from \rel\ electrons, where the electrons are accelerated 
by the \ultrarel\ but decelerating fireball shock. 
A further assumption that is generally made
is that the electrons obtain a power-law distribution with some minimum Lorentz factor that emerges from DSA 
\citep[e.g.,][]{Meszaros2002,Piran2004_05,Leventis2013}. Equally constraining, 
the shape of the 
electron distribution is often assumed to remain constant during the evolution of the blastwave as it slows in the circumstellar medium 
(CSM). 
As the PIC and \MC\ results we have discussed show, \ultrarel\ shock theory has difficulty supporting these widely used assumptions.

The ejecta moving relativistically from the central engine in a GRB will drive a forward shock (FS) into the circumstellar material (CSM) and a  reverse shock (RS) will propagate through the ejecta material 
\citep[e.g.,][ and references therein]{Piran2004_05}. Initially, the FS
will be \ultrarel\ but the RS may initially be \nonrel.

In the case of matter dominated ejecta with a density $\Delta$ times the density of the CSM, the Lorentz factor of the RS, $\gamRS$, in the rest frame of the ejecta, can be estimated as:
\begin{equation} \label{eq:gamRS}
     \gamRS \approx  \left\{
   \begin{array}{ll}
         1 + \frac{4}{7} \gamEj^2 \Delta^{-1},            &     
\gamEj^2  \ll \Delta  \\
          \frac{1}{\sqrt{2}} \gamEj^{1/2} \Delta^{-1/4},  &
         \gamEj^2   \gg \Delta \, ,
   \end{array}
     \right .
\end{equation}
where $\gamEj$ is the Lorentz factor of the ejecta 
\citep[see][]{Piran2004_05}. 
In the early
expansion stage, when the ejecta density is large and 
$\gamEj^2 \ll \Delta$, the reverse shock is \nonrel.
It will become mildly
relativistic when $\gamEj^2 \sim \Delta$.
While the energy in the RS
is much less than in the \ultrarel\ FS, we have shown 
(e.g., Figure~\ref{fig:fp_nonrel_to_rel}) that 
DSA is more efficient in \transrel\ shocks and the spectrum produced is harder than for \ultrarel\ shocks.

In the case of  GRBs driven by ejecta that is initially magnetically dominated \citep[e.g.,][]{lyutikov_blandford03,zhang_yan11},
a \transrel\ RS may form during the afterglow stage if  efficient magnetic dissipation occurs during the
prompt emission regime
\citep[e.g.,][]{zhang_yan11,BykovTreumann2011,BykovEtal2012}.

\begin{figure}
\epsscale{1.0}
\plotone{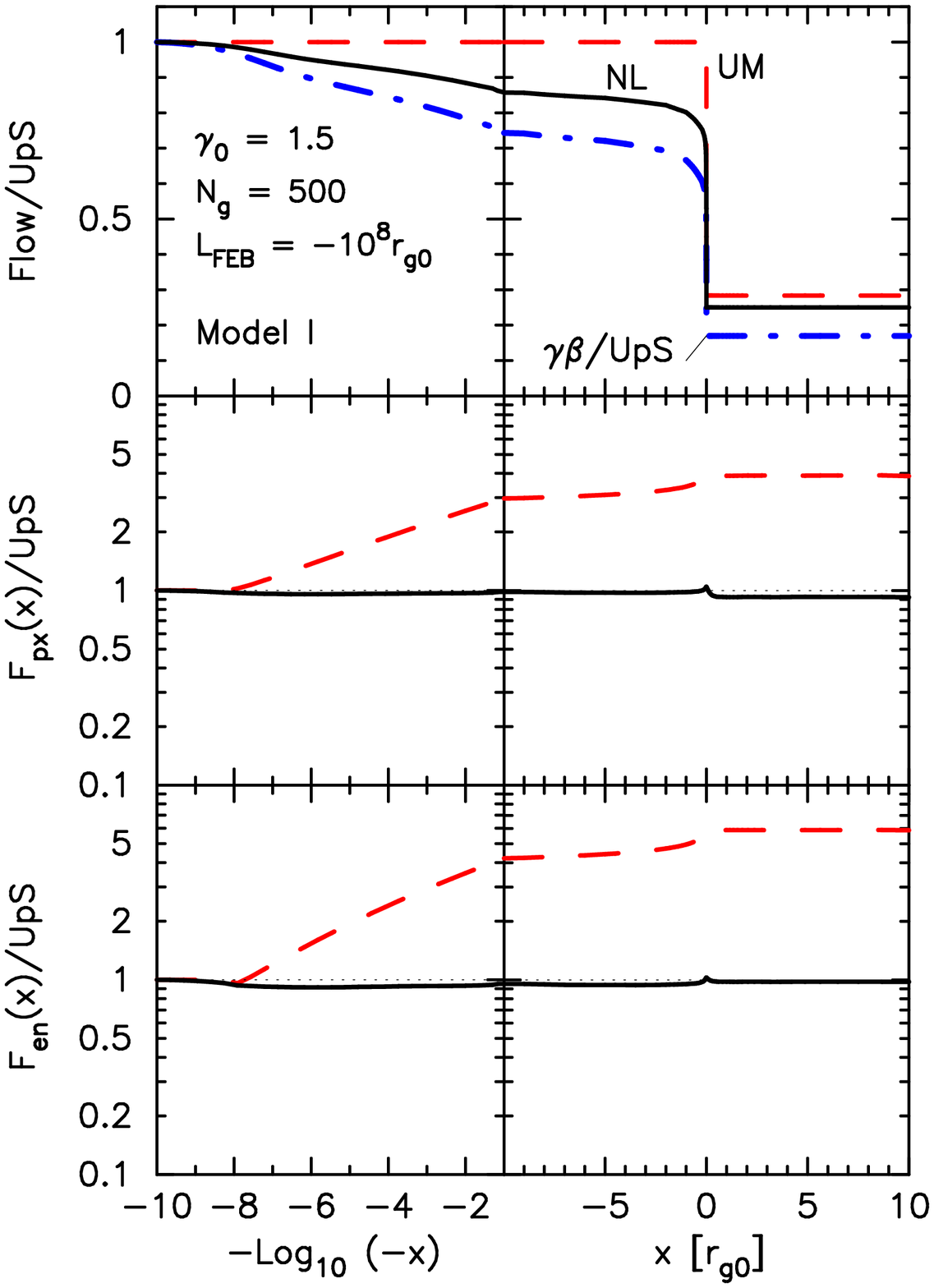} 
\caption{Structure for a $\gamZ=1.5$ shock with a FEB at $-10^8\,\rgZ$ (Model I). The figure format is the same as
Figure~\ref{fig:profile_gam1.5}. The \SC\ compression ratio is $\rNL=4.0\pm0.4$ compared to the \RH\ value $\rRH \simeq 3.53$. 
\label{fig:grid_eV18}}
\end{figure}

The main reason for considering a \transrel\ RS, rather than the \ultrarel\ FS,  that carries far more energy, as the source of the afterglow, are the conceptual difficulties 
superluminal \ultrarel\ shocks have in diffusively injecting and accelerating particles to energies needed to model GRB afterglows
\citep*[e.g.,][]{PLM2009}. 
Whatever difficulties \ultrarel\ shocks have should be less for   \transrel\ shocks. Furthermore,  our parallel-shock approximation
is more applicable to subluminal \transrel\ shocks.

\subsection{\Transrel\ Supernovae and UHECRs}
Recently, mildly \rel\ type Ibc supernovae have been discovered 
\citep[e.g.,][]{Soderberg2010,Paragi2010,Chakraborti2011} with 
inferred blast-wave speeds 
such that $\gamZ \betaZ \sim 1$.
These  speeds lie between normal supernova remnants (SNRs)
with $10^{-3} \lsim \gamZ \betaZ \lsim\, 0.1$ and GRB jets or
fireballs with 
$\gamZ \betaZ \gsim 5$ and suggest there
may be a continuum of shock speeds from the explosions of massive stars over the full range from \nonrel\ to \ultrarel\ 
\citep[see, for example,][]{LazzatiEtal2012}.

\begin{figure}
\epsscale{1.0}
\plotone{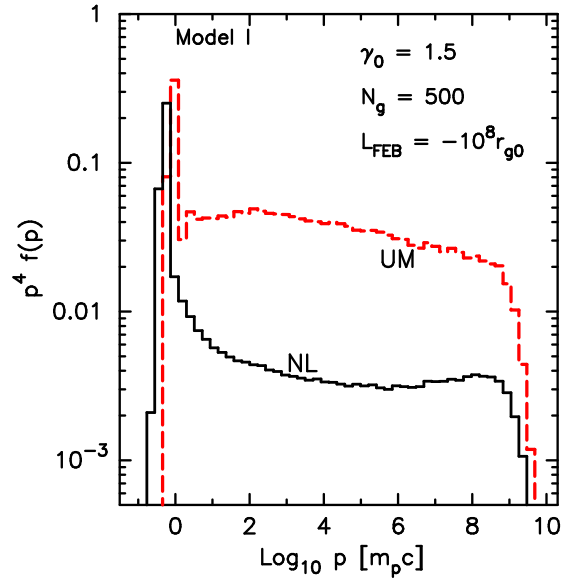} 
\caption{Comparison of UM downstream spectrum (dashed, red curve) for our Model I with $\Lfeb = -10^8\,\rgZ$ against the NL spectrum (solid, black curve).
\label{fig:fp_eV18}}
\end{figure}

Our \transrel\ results   are important for the interpretation of emission from shock accelerated particles in this particular class of mildly \rel\ type Ibc SNe.
We showed in Figure~\ref{fig:fp_gam1.5} that phase-space spectra $f(p) \propto p^{-4}$ can be expected in a transitional zone around $\gamZ \sim 1.5$ ($\gamZ \betaZ \sim 1$).
Perhaps more importantly however, all \rel\ blast-wave shocks will 
slow as they sweep up material and, if DSA is working as modeled in Figure~\ref{fig:fp_nonrel_to_rel},  \rel\ shocks will surely pass through a transitional phase. 

\citet{Budnik2008} have  suggested that galactic \transrel\ SNe may produce cosmic rays (CRs) with energies in the range
$\sim 10^{15}$ to $\sim 10^{18}$\,eV. 
An even more dramatic assertion by \citet{Chakraborti2011} is that the sub-population of type Ibc SNe such as SN 2009bb may be able to produce UHECRs to the \GZK\ (GZK) limit.

In our steady-state simulation, for a given set of shock parameters including $\etamfp$ and $N_g$, the maximum energy a shock can produce, $\Emax$, is proportional to  the distance to the FEB, $\Lfeb$.
Of course, in our plane-shock model, the FEB mimics the effects of a finite shock where $\Lfeb$ corresponds to the scale set by the shock radius or system size, an approximation that has been used to good effect in models of young \nonrel\ SNRs \citep[e.g.,][]{LSENP2013}.
In Figures~\ref{fig:grid_eV18} and \ref{fig:fp_eV18} (Model I),
we show results for a $\gamZ=1.5$ shock 
where $\Lfeb=-10^8\rgZ \sim 0.25$\,pc and $B_0 = 300$\,\muG. All other parameters are
the same as those used for Model C shown in 
Figures~\ref{fig:profile_gam1.5} and \ref{fig:fp_gam1.5}.
$\Lfeb$ and $B_0$ have been chosen to obtain $\Emax \sim 10^{18}$\,eV before the turnover from CRs escaping at the FEB.
The large $B_0$ assumes that this $\gamZ=1.5$ shock can produce strong \MFA\ or that it exists in a region with large ambient fields.
While the evidence supporting strong MFA in \nonrel\ shocks is convincing \citep[e.g.,][]{CassamEtal2007}, little MFA is expected in \ultrarel\ shocks. However, since MFA is driven by CR production, we expect the spectral transition seen in 
Figure~\ref{fig:fp_nonrel_to_rel} to be accompanied by MFA as the efficiency of DSA increases and the shock speed slows.

As indicated by the dot-dashed (blue) curve in the top panels of 
Figure~\ref{fig:grid_eV18}, the shock precursor
is noticeably smoothed over the entire region from $\Lfeb$ to the subshock.
In Figure~\ref{fig:fp_eV18}, the concave shape of the 
NL spectrum (solid, black curve) is  
obvious but, to a fair approximation,  the spectrum is harder than the UM one and $f(p) \propto p^{-4}$ over a wide momentum range.
Of course, we have made the assumption that Bohm diffusion, with $B_0 \sim 300$\,\muG, occurs over the scale indicated 
in Figure~\ref{fig:grid_eV18}
\citep[e.g.,][]{MN2006}.
If this in fact happens, \transrel\ shocks might 
be able to produce CRs to $\sim 10^{18}$\,eV.

Another possible source of UHECRs involving DSA in \transrel\ shocks are  jets in active galactic nuclei (AGN) 
\citep[see, for example,][ for reviews]{Kotera_Olinto2011,ABPW2012}.
Shock speeds of $\betaZ \sim 0.3$ have been inferred from  
multi-waveband observations in a number of radio hot spots \citep[e.g.,][]{Meisenheimer1989} and \citet{RomeroEtal1996} suggest that the \transrel\ jet in \CenA\ (NGC 5128), the closest AGN, may accelerate CRs to $\Emax >10^{21}$\,eV. For this $\Emax$, 
a shock size of $\sim 2$\,kpc and a field of $\sim 10$\,\muG\ were assumed.
\citet{Lemoine_Waxman2009} 
established that a jet magnetic luminosity of $L_B \gtrsim 10^{45} Z^{-2}$ erg s$^{-1}$ 
is needed in order to accelerate particles of charge number
$Z$ up to 100 EeV.
Since $L_B$ for \CenA\ is estimated to be 
$\gsim 10^{43}$\,erg s$^{-1}$, the production of CRs to $\sim 10^{21}$\,eV may be possible.

\newlistroman

\section{Conclusions} \label{sec:Conc}
If collisionless shocks accelerate particles efficiently
the shock structure must be calculated \SCly\ with the acceleration and particle escape
to conserve energy and momentum.
We have presented a \MC\ model of \NL\ DSA that, once the scattering properties are defined, 
determines the thermal injection and 
shock structure in plane-parallel, steady-state shocks of arbitrary Lorentz factor.
Our injection model, parameterized diffusion, 
and plane-parallel, steady-state approximations are
restrictive when compared to PIC simulations and some
\MC\ calculations that trace particle orbits.
However, apart from PIC simulations and the 
preliminary work of \citet{ED2002}, 
they are the first to model DSA in NL \rel\ shocks.
\NLin\ effects may be critical for understanding GRB afterglows, blazars \citep[e.g.,][]{SteckerEtal2007}, radio jets 
\citep[e.g.,][]{WykesEtal2013}, and DSA in a newly discovered class  of mildly \rel\ type Ibc SNe
\citep[e.g.,][]{Budnik2008,Soderberg2010}.
Next is a list of our major findings.

\listromanDE 
In terms of the backreaction of CRs on the shock structure,
 NL DSA works identically in \nonrel\ and 
\rel\ shocks, since the effects we consider depend only on energy and momentum  conservation.
Quantitative differences result from the fact that \rel\ shocks tend to have a lower intrinsic compression ratio than \nonrel\ shocks and because particle distributions are highly anisotropic in \rel\ shocks. Besides making the analysis of NL DSA more difficult, this anisotropy makes \rel\ shocks much more 
sensitive to assumptions made regarding particle diffusion and, unlike in \nonrel\ shocks, these assumptions can strongly influence the acceleration process.
We do not specify the production mechanism of self-generated magnetic turbulence
here but rather assume that Bohm diffusion applies.
There are fundamental problems in how, or even if, this
turbulence is created that depend on shock speed and obliquity
\citep[see, for example,][ and references therein]{PPL2013}.

\listromanDE
There is a smooth transition between the 
concave upward spectral shape
long established for \nonrel\ shocks undergoing efficient DSA and the 
soft
power-law spectrum predicted for fully \rel\ shocks 
(i.e., Figure~\ref{fig:fp_nonrel_to_rel}). 
For the parameters used here, this transition occurs around $\gamZ=1.5$, a speed recently associated with a subclass of type Ibc SNe \citep[e.g.,][]{Soderberg2010}. 
An increase in MFA can also be expected as shocks slow and cross this transition zone.

\listromanDE
Even though NL effects from particle acceleration are weaker in fully \rel\ shocks than in \nonrel\ ones, they nevertheless have a substantial effect on the accelerated particle spectrum.
Our NL results for $\gamZ=10$, in the fine scattering limit, show  the 
normalization of the energetic particle distribution dropping by a factor of about 5 compared to the  UM shock 
(see Figure~\ref{fig:fp_gam10}).
While these specific results depend quantitatively on the details of our model, basic energy and momentum conservation demands similar backreaction effects if DSA is efficient.

\listromanDE If \LAS\ is assumed in \rel\ shocks, particles can gain a much larger amount of energy in a given shock crossing than for fine \PAS. 
While this suggests that these shocks can be extremely efficient accelerators with hard spectra,  NL shock smoothing reduces the resultant efficiency and softens the spectrum considerably 
(see Figures~\ref{fig:fp_gam10_Ng100} and \ref{fig:fp_Baring_Ng4}).

\listromanDE Not forgetting the fundamental differences between our \MC\ technique and PIC simulations, we have compared our results with those of \citet{SironiSpit2011} and shown that 
important similarities exist in the spectral shape, average shock structure, and overall acceleration efficiency determined by each technique for 
a quasi-parallel, subluminal shock 
(see Figures~\ref{fig:grid_gam15_FEB3} and \ref{fig:fp_Acc_FEB3}). 
Our results also suggest that short time-scale shock reformation effects can be adequately modeled with a steady-state simulation.

\listromanDE
The major reason for employing the \MC\ technique, when far more physically complete PIC simulations exist, is indicated by 
Figures~\ref{fig:grid_eV18} and \ref{fig:fp_eV18}. It has been suggested that \transrel\ shocks may be able to produce CRs to $10^{18}$\,eV and above. In order to model such acceleration, large spatial and time scales are required which remain well beyond PIC limits. To obtain $\Emax \sim 10^{18}$\,eV, we needed 
$\Lfeb = -10^8\,\rgZ$ for our $\gamZ=1.5$ shock, compared to  
$\Lfeb = -3\,\rgZ$ used for our $\gamZ=15$ comparison to 
the \citet{SironiSpit2011} results (Model G). After  adjusting for the different magnetic fields and shock speeds used in our length unit $\rgZ$, this is a factor of $\sim 2.5\xx{5}$ difference
in real units, a factor easily accommodated by the \MC\ technique.

\newlistroman

The most important approximations made by the \MC\ technique are: \\
\listromanDE 
The particle scattering is parameterized using 
$\etamfp$ (equation~\ref{eq:mfp}) and $N_g$ (equation~\ref{eq:Tmax}).
For this study, we only show results for Bohm diffusion, i.e., $\etamfp=1$, although our plane-parallel results are independent of $\etamfp$ if the one length scale in the problem, the FEB, is scaled in particle gyroradii.
Our scattering parameterization has no spatial dependence and does not include \MFA, a process almost certain to be important in strong, \nonrel\ shocks 
\citep[e.g.,][]{Bell2004,VBE2008,ESPB2012}, and one that has been invoked to explain UHECR production in \transrel\ shocks \citep[e.g.,][]{Chakraborti2011}.
We note that the high acceleration efficiencies we find, and the related concave spectral shapes, may be less extreme in oblique shocks where values of $\etamfp > 1$. 
Difficulties in producing strong turbulence in \ultrarel, superluminal shocks may also result in larger effective $\etamfp$'s, reducing NL effects from DSA and resulting in lower cutoff energies in any given shock environment.
In the critical \transrel, subluminal regime, however, effects from shock obliquity are less restrictive and strong turbulence is more likely to be produced, making our Bohm diffusion approximation more plausible.

\listromanDE
The damping of magnetic turbulence isn't included.  
The strong NL effects we see will be lessened if 
the shock precursor is heated as energy is transferred from the turbulence to the background plasma.

\listromanDE
We only model plane, parallel shocks while \rel\ shocks
are expected to be oblique. Based on PIC simulation results 
\citep[e.g.,][]{SironiSpit2011}, the consequences of this restriction should be minor as long as the shock remains subluminal. However, important \rel\ shock applications, such as the termination shock in a pulsar wind nebula, are likely to be strongly superluminal. While  oblique, \rel, \TP\ shocks have been studied with \MC\ techniques 
\citep[e.g.,][]{ED2004,SummerlinBaring2012}, NL effects have not yet been modeled. 

\listromanDE
We have only modeled protons while applications involving radiation signatures of \rel\ shocks require electrons to be accelerated \SCly\ with  protons. Electrons can be included in the \MC\ model 
\citep[e.g.,][]{BaringEtal99}  but, unlike \SC\ PIC results, this requires additional assumptions and free parameters.
Electrons will be considered in future work. 

\listromanDE
Our \MC\ model is steady-state while most astrophysical shocks are in evolving systems. 
However, as suggested by 
Figure~\ref{fig:fp_nonrel_to_rel}, an evolving system can be approximated with several steady-state shocks, each differing in
parameters such as radius, speed, ambient density, and magnetic field strength.  

Diffusive shock acceleration is the most intensively investigated acceleration mechanism in astrophysics for good reason.
Collisionless shocks are common, some \nonrel\ shocks are known 
from {\it in situ} spacecraft observations to be efficient accelerators with self-generated  magnetic turbulence, and detailed predictions from \nonrel\
DSA theory provide excellent fits to some nonthermal sources
\citep*[see][ for a review]{LeeEtal2012}.
From its introduction in 1976-78, the theory of DSA developed rapidly, benefiting from the combined input from direct spacecraft observations of heliospheric shocks, analytic studies, and computer simulations.
The extension of DSA theory to \ultrarel\ shocks is a natural progression but definitive results have been slowed by the inherent difficulty  associated with magnetic turbulence generation and particle transport in \rel\ plasmas, and by the fact that no \rel\ shocks are directly observable by spacecraft.

What is certain is that there must be a smooth transition between efficient, \nonrel\ shocks, where observational evidence
suggests that DSA produces strong MFA in the shock precursor, and \ultrarel\ shocks,
which may be far less efficient accelerators and where self-generated turbulence may be weak or absent.
The \MC\ model we present here provides important information on the critical \transrel\ regime, albeit for a simple parameterized scattering model. More physically motivated diffusion has been included in \nonrel\ versions of the code 
\citep[e.g.,][]{VBE2008,VBE2009} and, in principle, similar generalizations can be made to the \rel\ code.

\acknowledgments The authors wish to thank Davide Lazzati, Herman Lee, Sergei Osipov,
Steve Reynolds, Lorenzo Sironi,  and Anatoli Spitkovsky for helpful discussions. 
D.C.E. and D.C.W. acknowledge support from NASA
grant NNX11AE03G. A.M.B. was partially supported by the RAS Presidium Programm and the RAS OFN Programm n17.
The authors also acknowledge helpful comments from the referee.

\bibliographystyle{aa} 
\bibliography{bib_DCE}

\begin{thebibliography}{85}
\expandafter\ifx\csname natexlab\endcsname\relax\def\natexlab#1{#1}\fi

\bibitem[{{Aharonian} {et~al.}(2012){Aharonian}, {Bykov}, {Parizot}, {Ptuskin},
  \& {Watson}}]{ABPW2012}
{Aharonian}, F., {Bykov}, A., {Parizot}, E., {Ptuskin}, V., \& {Watson}, A.
  2012, \ssr, 166, 97

\bibitem[{{Axford} {et~al.}(1977){Axford}, {Leer}, \& {Skadron}}]{ALS77}
{Axford}, W.~I., {Leer}, E., \& {Skadron}, G. 1977, Proc. 15th ICRC(Plovdiv),
  11, 132

\bibitem[{{Baring}(1999)}]{Baring1999}
{Baring}, M. 1999, in International Cosmic Ray Conference, Vol.~4,
  International Cosmic Ray Conference, 5

\bibitem[{{Baring}(2004)}]{Baring2004}
{Baring}, M.~G. 2004, Nuclear Physics B Proceedings Supplements, 136, 198

\bibitem[{Baring {et~al.}(1999)Baring, Ellison, Reynolds, Grenier, \&
  Goret}]{BaringEtal99}
Baring, M.~G., Ellison, D.~C., Reynolds, S.~P., Grenier, I.~A., \& Goret, P.
  1999, ApJ, 513, 311

\bibitem[{{Baring} {et~al.}(1997){Baring}, {Ogilvie}, {Ellison}, \&
  {Forsyth}}]{BOEF97}
{Baring}, M.~G., {Ogilvie}, K.~W., {Ellison}, D.~C., \& {Forsyth}, R.~J. 1997,
  \apj, 476, 889

\bibitem[{{Bednarz} \& {Ostrowski}(1998)}]{BO98}
{Bednarz}, J. \& {Ostrowski}, M. 1998, Physical Review Letters, 80, 3911

\bibitem[{{Bell}(1978)}]{Bell78a}
{Bell}, A.~R. 1978, \mnras, 182, 147

\bibitem[{{Bell}(2004)}]{Bell2004}
{Bell}, A.~R. 2004, \mnras, 353, 550

\bibitem[{{Bell}(2005)}]{Bell2005}
{Bell}, A.~R. 2005, \mnras, 358, 181

\bibitem[{Berezhko \& Ellison(1999)}]{BE99}
Berezhko, E.~G. \& Ellison, D.~C. 1999, ApJ, 526, 385

\bibitem[{{Blandford} \& {Eichler}(1987)}]{BE87}
{Blandford}, R. \& {Eichler}, D. 1987, Physics Reports, 154, 1

\bibitem[{{Blandford} \& {Ostriker}(1978)}]{BO78}
{Blandford}, R.~D. \& {Ostriker}, J.~P. 1978, \apjl, 221, L29

\bibitem[{{Blasi} \& {Vietri}(2005)}]{BV2005}
{Blasi}, P. \& {Vietri}, M. 2005, \apj, 626, 877

\bibitem[{{Budnik} {et~al.}(2008){Budnik}, {Katz}, {MacFadyen}, \&
  {Waxman}}]{Budnik2008}
{Budnik}, R., {Katz}, B., {MacFadyen}, A., \& {Waxman}, E. 2008, \apj, 673, 928

\bibitem[{{Bykov} {et~al.}(2012){Bykov}, {Gehrels}, {Krawczynski}, {Lemoine},
  {Pelletier}, \& {Pohl}}]{BykovEtal2012}
{Bykov}, A., {Gehrels}, N., {Krawczynski}, H., {et~al.} 2012, \ssr, 173, 309

\bibitem[{{Bykov} \& {Treumann}(2011)}]{BykovTreumann2011}
{Bykov}, A.~M. \& {Treumann}, R.~A. 2011, \aapr, 19, 42

\bibitem[{{Caprioli} {et~al.}(2009){Caprioli}, {Blasi}, {Amato}, \&
  {Vietri}}]{CBAV2009}
{Caprioli}, D., {Blasi}, P., {Amato}, E., \& {Vietri}, M. 2009, \mnras, 395,
  895

\bibitem[{{Cassam-Chena{\"i}} {et~al.}(2007){Cassam-Chena{\"i}}, {Hughes},
  {Ballet}, \& {Decourchelle}}]{CassamEtal2007}
{Cassam-Chena{\"i}}, G., {Hughes}, J.~P., {Ballet}, J., \& {Decourchelle}, A.
  2007, \apj, 665, 315

\bibitem[{{Chakraborti} {et~al.}(2011){Chakraborti}, {Ray}, {Soderberg},
  {Loeb}, \& {Chandra}}]{Chakraborti2011}
{Chakraborti}, S., {Ray}, A., {Soderberg}, A.~M., {Loeb}, A., \& {Chandra}, P.
  2011, Nature Communications, 2

\bibitem[{{Double} {et~al.}(2004){Double}, {Baring}, {Jones}, \&
  {Ellison}}]{DoubleEtal2004}
{Double}, G.~P., {Baring}, M.~G., {Jones}, F.~C., \& {Ellison}, D.~C. 2004,
  \apj, 600, 485

\bibitem[{{Drury}(1983)}]{Drury83}
{Drury}, L.~O. 1983, Reports of Progress in Physics, 46, 973

\bibitem[{{Ellison}(1985)}]{Ellison85}
{Ellison}, D.~C. 1985, \jgr, 90, 29

\bibitem[{{Ellison} {et~al.}(1996){Ellison}, {Baring}, \& {Jones}}]{EBJ96}
{Ellison}, D.~C., {Baring}, M.~G., \& {Jones}, F.~C. 1996, \apj, 473, 1029

\bibitem[{{Ellison} \& {Double}(2002)}]{ED2002}
{Ellison}, D.~C. \& {Double}, G.~P. 2002, Astroparticle Physics, 18, 213

\bibitem[{{Ellison} \& {Double}(2004)}]{ED2004}
{Ellison}, D.~C. \& {Double}, G.~P. 2004, Astroparticle Physics, 22, 323

\bibitem[{{Ellison} \& {Eichler}(1984)}]{EE84}
{Ellison}, D.~C. \& {Eichler}, D. 1984, \apj, 286, 691

\bibitem[{{Ellison} {et~al.}(1993){Ellison}, {Giacalone}, {Burgess}, \&
  {Schwartz}}]{EGBS93}
{Ellison}, D.~C., {Giacalone}, J., {Burgess}, D., \& {Schwartz}, S.~J. 1993,
  \jgr, 98, 21085

\bibitem[{{Ellison} {et~al.}(1990{\natexlab{a}}){Ellison}, {Jones}, \&
  {Reynolds}}]{EJR90}
{Ellison}, D.~C., {Jones}, F.~C., \& {Reynolds}, S.~P. 1990{\natexlab{a}},
  \apj, 360, 702

\bibitem[{{Ellison} {et~al.}(1990{\natexlab{b}}){Ellison}, {Moebius}, \&
  {Paschmann}}]{EMP90}
{Ellison}, D.~C., {Moebius}, E., \& {Paschmann}, G. 1990{\natexlab{b}}, \apj,
  352, 376

\bibitem[{{Ellison} \& {Reynolds}(1991)}]{ER91}
{Ellison}, D.~C. \& {Reynolds}, S.~P. 1991, \apj, 378, 214

\bibitem[{{Ellison} {et~al.}(2012){Ellison}, {Slane}, {Patnaude}, \&
  {Bykov}}]{ESPB2012}
{Ellison}, D.~C., {Slane}, P., {Patnaude}, D.~J., \& {Bykov}, A.~M. 2012, \apj,
  744, 39

\bibitem[{{Giacalone} \& {Ellison}(2000)}]{GE2000}
{Giacalone}, J. \& {Ellison}, D.~C. 2000, \jgr, 105, 12541

\bibitem[{{Hoshino} {et~al.}(1992){Hoshino}, {Arons}, {Gallant}, \&
  {Langdon}}]{Hoshino92}
{Hoshino}, M., {Arons}, J., {Gallant}, Y.~A., \& {Langdon}, A.~B. 1992, \apj,
  390, 454

\bibitem[{{Jokipii}(1971)}]{Jokipi1971}
{Jokipii}, J.~R. 1971, Reviews of Geophysics and Space Physics, 9, 27

\bibitem[{{Jones} \& {Ellison}(1991)}]{JE91}
{Jones}, F.~C. \& {Ellison}, D.~C. 1991, Space Science Reviews, 58, 259

\bibitem[{{Jones} {et~al.}(1998){Jones}, {Jokipii}, \& {Baring}}]{JJB98}
{Jones}, F.~C., {Jokipii}, J.~R., \& {Baring}, M.~G. 1998, \apj, 509, 238

\bibitem[{{Kato}(2007)}]{Kato2007}
{Kato}, T.~N. 2007, \apj, 668, 974

\bibitem[{{Katz} {et~al.}(2010){Katz}, {M{\'e}sz{\'a}ros}, \&
  {Waxman}}]{KMW2010}
{Katz}, B., {M{\'e}sz{\'a}ros}, P., \& {Waxman}, E. 2010, \jcap, 10, 12

\bibitem[{{Keshet} {et~al.}(2009){Keshet}, {Katz}, {Spitkovsky}, \&
  {Waxman}}]{KeshetEtal2009}
{Keshet}, U., {Katz}, B., {Spitkovsky}, A., \& {Waxman}, E. 2009, \apjl, 693,
  L127

\bibitem[{{Keshet} \& {Waxman}(2005)}]{KW2005}
{Keshet}, U. \& {Waxman}, E. 2005, Physical Review Letters, 94, 111102

\bibitem[{{Kirk} {et~al.}(2000){Kirk}, {Guthmann}, {Gallant}, \&
  {Achterberg}}]{KirkEtal2000}
{Kirk}, J.~G., {Guthmann}, A.~W., {Gallant}, Y.~A., \& {Achterberg}, A. 2000,
  \apj, 542, 235

\bibitem[{{Kotera} \& {Olinto}(2011)}]{Kotera_Olinto2011}
{Kotera}, K. \& {Olinto}, A.~V. 2011, \araa, 49, 119

\bibitem[{{Krymskii}(1977)}]{Kry77}
{Krymskii}, G.~F. 1977, Akademiia Nauk SSSR Doklady, 234, 1306

\bibitem[{{Lazzati} {et~al.}(2012){Lazzati}, {Morsony}, {Blackwell}, \&
  {Begelman}}]{LazzatiEtal2012}
{Lazzati}, D., {Morsony}, B.~J., {Blackwell}, C.~H., \& {Begelman}, M.~C. 2012,
  \apj, 750, 68

\bibitem[{{Lee} {et~al.}(2012){Lee}, {Mewaldt}, \& {Giacalone}}]{LeeEtal2012}
{Lee}, M.~A., {Mewaldt}, R.~A., \& {Giacalone}, J. 2012, \ssr, 173, 247

\bibitem[{{Lee} {et~al.}(2013){Lee}, {Slane}, {Ellison}, {Nagataki}, \&
  {Patnaude}}]{LSENP2013}
{Lee}, S.-H., {Slane}, P.~O., {Ellison}, D.~C., {Nagataki}, S., \& {Patnaude},
  D.~J. 2013, \apj, 767, 20

\bibitem[{{Lemoine} \& {Pelletier}(2003)}]{LP2003}
{Lemoine}, M. \& {Pelletier}, G. 2003, \apjl, 589, L73

\bibitem[{{Lemoine} \& {Revenu}(2006)}]{LR2006}
{Lemoine}, M. \& {Revenu}, B. 2006, \mnras, 366, 635

\bibitem[{{Lemoine} \& {Waxman}(2009)}]{Lemoine_Waxman2009}
{Lemoine}, M. \& {Waxman}, E. 2009, \jcap, 11, 9

\bibitem[{{Leventis} {et~al.}(2013){Leventis}, {van der Horst}, {van Eerten},
  \& {Wijers}}]{Leventis2013}
{Leventis}, K., {van der Horst}, A.~J., {van Eerten}, H.~J., \& {Wijers},
  R.~A.~M.~J. 2013, \mnras, 431, 1026

\bibitem[{{Lyutikov} \& {Blandford}(2003)}]{lyutikov_blandford03}
{Lyutikov}, M. \& {Blandford}, R. 2003, ArXiv Astrophysics e-prints

\bibitem[{{Malkov} \& {Drury}(2001)}]{MD2001}
{Malkov}, M.~A. \& {Drury}, L. 2001, Reports of Progress in Physics, 64, 429

\bibitem[{{McKenzie} \& {Voelk}(1982)}]{MV82}
{McKenzie}, J.~F. \& {Voelk}, H.~J. 1982, \aap, 116, 191

\bibitem[{{Meisenheimer} {et~al.}(1989){Meisenheimer}, {Roser}, {Hiltner},
  {Yates}, {Longair}, {Chini}, \& {Perley}}]{Meisenheimer1989}
{Meisenheimer}, K., {Roser}, H.-J., {Hiltner}, P.~R., {et~al.} 1989, \aap, 219,
  63

\bibitem[{{Meli} {et~al.}(2008){Meli}, {Becker}, \& {Quenby}}]{Meli_Quenby2008}
{Meli}, A., {Becker}, J.~K., \& {Quenby}, J.~J. 2008, \aap, 492, 323

\bibitem[{{M{\'e}sz{\'a}ros}(2002)}]{Meszaros2002}
{M{\'e}sz{\'a}ros}, P. 2002, \araa, 40, 137

\bibitem[{{Milosavljevi{\'c}} \& {Nakar}(2006)}]{MN2006}
{Milosavljevi{\'c}}, M. \& {Nakar}, E. 2006, \apj, 651, 979

\bibitem[{{Niemiec} \& {Ostrowski}(2006)}]{NO2006}
{Niemiec}, J. \& {Ostrowski}, M. 2006, \apj, 641, 984

\bibitem[{{Niemiec} {et~al.}(2006){Niemiec}, {Ostrowski}, \&
  {Pohl}}]{NiemiecEtal2006}
{Niemiec}, J., {Ostrowski}, M., \& {Pohl}, M. 2006, \apj, 650, 1020

\bibitem[{{Nishikawa} {et~al.}(2007){Nishikawa}, {Hededal}, {Hardee},
  {Fishman}, {Kouveliotou}, \& {Mizuno}}]{NishikawaEtal2007}
{Nishikawa}, K.-I., {Hededal}, C.~B., {Hardee}, P.~E., {et~al.} 2007, \apss,
  307, 319

\bibitem[{{Ostrowski}(1988)}]{Ostrowski1988}
{Ostrowski}, M. 1988, \mnras, 233, 257

\bibitem[{{Ostrowski}(1991)}]{Ostrowski1991}
{Ostrowski}, M. 1991, \mnras, 249, 551

\bibitem[{{Ostrowski}(1993)}]{Ostrowski1993}
{Ostrowski}, M. 1993, \mnras, 264, 248

\bibitem[{{Paragi} {et~al.}(2010){Paragi}, {Taylor}, {Kouveliotou}, {Granot},
  {Ramirez-Ruiz}, {Bietenholz}, {van der Horst}, {Pidopryhora}, {van
  Langevelde}, {Garrett}, {Szomoru}, {Argo}, {Bourke}, \&
  {Paczy{\'n}ski}}]{Paragi2010}
{Paragi}, Z., {Taylor}, G.~B., {Kouveliotou}, C., {et~al.} 2010, \nat, 463, 516

\bibitem[{{Pelletier} {et~al.}(2006){Pelletier}, {Lemoine}, \&
  {Marcowith}}]{PLM2006}
{Pelletier}, G., {Lemoine}, M., \& {Marcowith}, A. 2006, \aap, 453, 181

\bibitem[{{Pelletier} {et~al.}(2009){Pelletier}, {Lemoine}, \&
  {Marcowith}}]{PLM2009}
{Pelletier}, G., {Lemoine}, M., \& {Marcowith}, A. 2009, \mnras, 393, 587

\bibitem[{{Piran}(2004)}]{Piran2004_05}
{Piran}, T. 2004, Reviews of Modern Physics, 76, 1143

\bibitem[{{Plotnikov} {et~al.}(2011){Plotnikov}, {Pelletier}, \&
  {Lemoine}}]{PPL2011}
{Plotnikov}, I., {Pelletier}, G., \& {Lemoine}, M. 2011, \aap, 532, A68

\bibitem[{{Plotnikov} {et~al.}(2013){Plotnikov}, {Pelletier}, \&
  {Lemoine}}]{PPL2013}
{Plotnikov}, I., {Pelletier}, G., \& {Lemoine}, M. 2013, \mnras, 430, 1280

\bibitem[{{Romero} {et~al.}(1996){Romero}, {Combi}, {Perez Bergliaffa}, \&
  {Anchordoqui}}]{RomeroEtal1996}
{Romero}, G.~E., {Combi}, J.~A., {Perez Bergliaffa}, S.~E., \& {Anchordoqui},
  L.~A. 1996, Astroparticle Physics, 5, 279

\bibitem[{{Sagi} \& {Nakar}(2012)}]{SagiNakar2012}
{Sagi}, E. \& {Nakar}, E. 2012, \apj, 749, 80

\bibitem[{{Schure} {et~al.}(2012){Schure}, {Bell}, {O'C Drury}, \&
  {Bykov}}]{SchureEtal2012}
{Schure}, K.~M., {Bell}, A.~R., {O'C Drury}, L., \& {Bykov}, A.~M. 2012, \ssr,
  173, 491

\bibitem[{{Sironi} \& {Spitkovsky}(2009)}]{SironiSpit2009}
{Sironi}, L. \& {Spitkovsky}, A. 2009, \apj, 698, 1523

\bibitem[{{Sironi} \& {Spitkovsky}(2011)}]{SironiSpit2011}
{Sironi}, L. \& {Spitkovsky}, A. 2011, \apj, 726, 75

\bibitem[{{Soderberg} {et~al.}(2010){Soderberg}, {Chakraborti}, {Pignata},
  {Chevalier}, {Chandra}, {Ray}, {Wieringa}, {Copete}, {Chaplin},
  {Connaughton}, {Barthelmy}, {Bietenholz}, {Chugai}, {Stritzinger}, {Hamuy},
  {Fransson}, {Fox}, {Levesque}, {Grindlay}, {Challis}, {Foley}, {Kirshner},
  {Milne}, \& {Torres}}]{Soderberg2010}
{Soderberg}, A.~M., {Chakraborti}, S., {Pignata}, G., {et~al.} 2010, \nat, 463,
  513

\bibitem[{{Spitkovsky}(2008)}]{Spitkovsky2008b}
{Spitkovsky}, A. 2008, \apjl, 682, L5

\bibitem[{{Stecker} {et~al.}(2007){Stecker}, {Baring}, \&
  {Summerlin}}]{SteckerEtal2007}
{Stecker}, F.~W., {Baring}, M.~G., \& {Summerlin}, E.~J. 2007, \apjl, 667, L29

\bibitem[{{Summerlin} \& {Baring}(2012)}]{SummerlinBaring2012}
{Summerlin}, E.~J. \& {Baring}, M.~G. 2012, \apj, 745, 63

\bibitem[{{Uchiyama} {et~al.}(2007){Uchiyama}, {Aharonian}, {Tanaka},
  {Takahashi}, \& {Maeda}}]{Uchiyama_J1713_2007}
{Uchiyama}, Y., {Aharonian}, F.~A., {Tanaka}, T., {Takahashi}, T., \& {Maeda},
  Y. 2007, \nat, 449, 576

\bibitem[{{Vladimirov} {et~al.}(2008){Vladimirov}, {Bykov}, \&
  {Ellison}}]{VBE2008}
{Vladimirov}, A.~E., {Bykov}, A.~M., \& {Ellison}, D.~C. 2008, \apj, 688, 1084

\bibitem[{{Vladimirov} {et~al.}(2009){Vladimirov}, {Bykov}, \&
  {Ellison}}]{VBE2009}
{Vladimirov}, A.~E., {Bykov}, A.~M., \& {Ellison}, D.~C. 2009, \apjl, 703, L29

\bibitem[{{Warren} {et~al.}(2005){Warren}, {Hughes}, {Badenes}, {Ghavamian},
  {McKee}, {Moffett}, {Plucinsky}, {Rakowski}, {Reynoso}, \&
  {Slane}}]{WarrenEtal2005}
{Warren}, J.~S., {Hughes}, J.~P., {Badenes}, C., {et~al.} 2005, \apj, 634, 376

\bibitem[{{Wykes} {et~al.}(2013){Wykes}, {Croston}, {Hardcastle}, {Eilek},
  {Biermann}, {Achterberg}, {Bray}, {Bicknell}, {Lazarian}, {Haverkorn},
  {Protheroe}, \& {Bromberg}}]{WykesEtal2013}
{Wykes}, S., {Croston}, J.~H., {Hardcastle}, M.~J., {et~al.} 2013, ArXiv
  e-prints

\bibitem[{{Zhang} \& {Yan}(2011)}]{zhang_yan11}
{Zhang}, B. \& {Yan}, H. 2011, \apj, 726, 90

\end{thebibliography}

\clearpage

\begin{table}
\begin{center}
\caption{Shock parameters.}
\label{tab:Input}
\vskip6pt
\begin{tabular}{crrrrrrrrrrrr}
\tableline
\tableline
\\
Model\tablenotemark{a}\tablenotetext{a}{For Model I, $B_0=300$\,\muG. All other models have $B_0=3$\,\muG. All models have $n_0=1$\,\pcc\ and $\etamfp=1$..}
&[$u_0$] \{$\betaZ$\} $\gamZ$ 
&$\Ng$
&$M_S$\tablenotemark{b}\tablenotetext{b}{We use the \nonrel\ definition for the sonic Mach number, $M_S^2 = \rho_0u_0^2/(\GamZ P_0)$. The far upstream ratio of specific heats $\GamZ=5/3$ in all cases.}
&$\rRH$\tablenotemark{c}\tablenotetext{c}{This is the \RH\ value as determined by equations~(\ref{eq:NumFlux})-(\ref{eq:EnFlux}) with no DSA.} 
&$\rNL$\tablenotemark{d}\tablenotetext{d}{This is the \SC\ total compression ratio with DSA.} 
&$|\Xsub|$
&$|\Lfeb|$
&$\Lfeb$ 
&$\EffDSA$\tablenotemark{e}\tablenotetext{e}{For the NL shock, $\EffDSA$ is the fraction of shock kinetic energy flux put into superthermal particles including those that leave at the upstream 
FEB. The number of significant figures suggests the level of accuracy.}
&$\qescEn$\tablenotemark{f}\tablenotetext{f}{For the NL shock, $\qescEn = \QescEn/\FenZ$ is the
fraction of shock kinetic energy flux that leaves the shock at the upstream FEB.}
\\
&[\kmps] & & & & &$\rgZ$ &$\rgZ$ &[pc] && \\
\\
\tableline
\\
A &[$2\xx{4}$] &20 &17 &3.96 &$12\pm1$ &0\tablenotemark{g}\tablenotetext{g}{A value of 0 means the \SC\ shock profile was found without using equation~(\ref{eq:Xsub}).}
&$3\xx{4}$ &$6\xx{-4}$ &$0.93$ &$0.5$ \\
B &\{0.2\} &60  &16 &3.93 &$11\pm1$ &0 &$1\xx{4}$ &$7\xx{-4}$ 
&$0.9$ &$0.45$ \\
C &1.5 &500 &$1900$ &3.53 &$4.0\pm0.4$ &0 &$1\xx{4}$ &$2.5\xx{-3}$ &$0.55$ &$0.02$ \\
D &10 &2000 &$2\xx{3}$ &3.02 &3.02 &$3$ &$1000$ &$3.4\xx{-4}$ 
&$0.3$ &$<0.01$ \\
E &30 &$10^4$ &$2\xx{3}$ &3.00 &3.00 &$9$ &$100$ &$3.4\xx{-5}$ 
&$0.35$ &$<0.01$ \\
F &10 &$100$ &$2\xx{3}$ &3.02 &3.02 &$1.75$ &$1000$ &$3.4\xx{-4}$ &$0.5$ &$0.01$ \\
G &15 &2000 &8 &3.00 &3.00 &4 &3 &$1\xx{-6}$ &0.3 &$<0.1$ \\
H &10.05 &4 &1800 &3.02 &3.02 &3.2 &$1\xx{5}$ &$3.4\xx{-2}$ &0.8 
&$<0.1$ \\
I &1.5 &500 &1900 &3.53 &$4.0\pm0.4$ &0 &$1\xx{8}$ &0.25 &0.6 &0.01 \\
\tableline
\tableline
\end{tabular}
\end{center}
\end{table}

\end{document}